\documentclass[aps,twocolumn,showpacs,superscriptaddress,groupedaddress,nofootinbib]{revtex4}  

\usepackage{graphicx}  
\usepackage{dcolumn}   
\usepackage{bm}        
\usepackage{amssymb}   
\usepackage{amsmath}   
\usepackage{multirow}
\usepackage{hyperref}
\usepackage{braket}
\usepackage{subfigure}
\usepackage{array}
\usepackage[math]{cellspace}
\usepackage[utf8]{inputenc}
\usepackage{cleveref}

\usepackage{color}
\usepackage[normalem]{ulem}

\newcommand{\simas}[1]{\raisebox{-.1ex}{
            $\stackrel{\small{#1}}{\sim}$}}

\newcommand{\ba}{\begin{array}}
\newcommand{\ea}{\end{array}}

\newcommand{\Dslash}{\relax{\kern+.25em / \kern-.70em D}}

\newcommand{\Real}{\relax{\mathsf{\Gamma\kern-.35em R}}}
\newcommand{\Int}{\relax{\mathsf{Z\kern-.40em Z}}}

\newcommand{\gbar}{\kern1pt\overline{\kern-1pt g\kern-0pt}\kern1pt}
\newcommand{\mbar}{\kern2pt\overline{\kern-1pt m\kern-1pt}\kern1pt}
\newcommand{\obar}[1]{\kern3pt\overline{\kern-2pt #1\kern-0pt}\kern1pt}

\newcommand{\abar}{\kern1pt\overline{\kern-1pt a\kern-0.5pt}\kern1pt}

\begin{document}

\hspace{5.2in} \mbox{IFIC/21-22}

\title{Topological sampling through windings}

\date{\today}

\author{David Albandea}
\affiliation{IFIC (CSIC-UVEG), Edificio Institutos Investigaci\'on, 
Apt.\ 22085, E-46071 Valencia, Spain}
\author{Pilar Hern\'andez}
\affiliation{IFIC (CSIC-UVEG), Edificio Institutos Investigaci\'on, 
Apt.\ 22085, E-46071 Valencia, Spain}
\author{Alberto Ramos}
\affiliation{IFIC (CSIC-UVEG), Edificio Institutos Investigaci\'on, 
Apt.\ 22085, E-46071 Valencia, Spain}
\author{Fernando Romero-L\'opez}
\affiliation{IFIC (CSIC-UVEG), Edificio Institutos Investigaci\'on, 
Apt.\ 22085, E-46071 Valencia, Spain}

\preprint{IFIC/21-22}

\begin{abstract}
	We propose a modification of the Hybrid Monte Carlo (HMC) algorithm that overcomes the topological freezing of a two-dimensional $U(1)$ gauge theory with and without fermion content. This algorithm includes reversible jumps between topological sectors---winding steps---combined with standard HMC steps. The full algorithm is referred to as winding HMC (wHMC), and it shows an improved behaviour of the autocorrelation time towards the continuum limit. We find excellent agreement between the wHMC estimates of the plaquette and topological susceptibility and the analytical predictions in the $U(1)$ pure gauge theory, which are known even at finite $\beta$. We also study the expectation values in fixed topological sectors using both HMC and wHMC, with and without fermions. Even when topology is frozen in HMC---leading to significant deviations in topological as well as non-topological quantities---the two algorithms agree on the fixed-topology averages. Finally, we briefly compare the wHMC algorithm results to those obtained with master-field simulations of size $L\sim 8 \times 10^3$.
\end{abstract}
\maketitle

\section{Introduction}
Standard algorithms for lattice QCD are well-known to suffer from topology freezing \cite{Alles:1996vn,DelDebbio:2002xa,DelDebbio:2004xh,Schaefer:2010hu}. Near the continuum limit, distinct topological sectors are poorly sampled due to the large energy barriers separating them, leading to exponentially increasing autocorrelation times as the continuum limit is approached in a finite volume. This problem has received a lot of attention, and several algorithmic strategies have been proposed over the years~\cite{Marinari:1992qd,Luscher:2011kk,Laio:2015era,Hasenbusch:2017unr,Bonanno:2020hht,Cossu:2021bgn}, but there is no fully satisfactory
solution.

In this paper, we study a modification of the Hybrid Monte Carlo (HMC) algorithm, named winding HMC (wHMC), that incorporates Metropolis--Hastings steps \cite{Hastings1970MonteCS} with tailored reversible jumps between topological sectors. The idea is similar to an old attempt under the name of instanton hit~\cite{Fucito:1983qg,Dilger:1994ma}. We will test our algorithm in the $U(1)$ gauge theory in 2D with and without fermion content. This model has been recently used as benchmark in machine-learned flow-based sampling algorithms~\cite{Kanwar:2020xzo,Albergo:2021vyo}, as well as in tensor network approaches~\cite{Funcke:2019zna,Butt:2019uul} (see Ref.~\cite{Banuls:2019bmf} for a review).

We first test the algorithm in a compact $U(1)$ pure gauge theory, which suffers
from topology freezing, but is solvable. Exact results on topological and
non-topological observables exist in the literature for the lattice
regularization~\cite{Kovacs_1995,Bonati:2019ylr,Bonati:2019olo}, i.e., at finite lattice spacing. Therefore, we can accurately test the approach to the continuum limit of the topological susceptibility and the plaquette. We then include two degenerate flavours of Wilson fermions and study the pion mass dependence of the topological susceptibility. In both cases we compare the scaling of the autocorrelation time with that of the standard HMC.

It is a general belief that algorithms with topology freezing do nevertheless well in observables computed at fixed topology---failing only in the weights of the different sectors. We can test this hypothesis accurately in this model by comparing the plaquette (in the pure gauge) and the pion mass (in the fermionic case) at fixed topology with the exact results or between the two algorithms, HMC and wHMC. 

Finally, since topology freezing can be circunvented altogether by working with very large physical volumes and taking local averages~\cite{Luscher:2017cjh,Giusti:2018cmp,Francis:2019muy}, we also compare our topology-sampling algorithm with HMC in a very large lattice of size $V=8192^2$. 

We comment on the prospects to extend the wHMC algorithm to other gauge theories and higher dimensions in the outlook section.

\section{Analytical results}

The Schwinger model~\cite{PhysRev.128.2425} is a $U(1)$ gauge theory with one or more massless fermions. It is a solvable quantum field theory that shares many properties with Yang--Mills in four dimensions \cite{Coleman:1975pw}. In particular, Euclidean gauge configurations can be classified according to their topological charge
\begin{eqnarray}
\nu = {1 \over 2 \pi} \int d^2 x  ~\epsilon^{\mu\nu} F_{\mu\nu}  \subset \mathbb{Z}, 
\end{eqnarray}
and there is a mass gap. The spectrum contains a free boson that can be interpreted as the singlet pseudoscalar meson, $\eta'$, with mass
\begin{eqnarray}
m_{\eta'}^2={N_f e^2 \over \pi} \equiv {N_f\over \pi \beta},
\end{eqnarray}
where $N_f$ is the number of degenerate flavours. 

Interestingly, the Witten-Veneziano formula is exact in the Schwinger model \cite{Seiler:1987ig,Giusti:2001xh}, 
\begin{eqnarray}
\chi_t|_{\rm quenched} \equiv \chi^q_t=
 {F_{\eta'}^2 m_{\eta'}^2\over 2 N_f} = {e^2 \over 4 \pi^2} = {1\over 4 \pi^2 \beta},
 \label{eq:xit}
\end{eqnarray}
where $F_{\eta'}$ is the decay constant, $F_{\eta'}= 1/\sqrt{2\pi}$, and the quenched topological susceptibility $\chi_{t}^{q}$ can be obtained in the pure gauge $U(1)$ theory in 2D. 

Since this theory can be solved on the lattice and in a finite volume
\cite{Kovacs_1995,Bonati:2019ylr,Bonati:2019olo}, it is therefore a good starting test-bed for Monte Carlo (MC) algorithms. 

\subsection{Compact $U(1)$ in 2D}

The Wilson lattice formulation of the theory is
\begin{eqnarray}
Z = \int \prod_{l} d U_l ~e^{-S_p[U]} \equiv \int \prod_{l} d U_l  ~e^{{\beta\over 2} \sum_p U_p+U_p^\dagger},
\end{eqnarray}
where $U_l$ and $U_p$ are the standard link and $1 \times 1$ Wilson loop,
respectively. We use periodic boundary conditions. Note that $\beta = 1/e^2$ is
dimensionful, but all dimensionful quantities are assumed in lattice units in
the following. Therefore, as we approach the continuum limit, $\beta \sim
a^{-2}$.

We will be considering  the plaquette and the topological susceptibility:
\begin{eqnarray}
P \equiv {\langle \sum_p {\rm Re}[U_p] \rangle\over V}, ~~~\chi_t \equiv  {\langle Q^2\rangle\over V},
\end{eqnarray}
where the lattice definition of topological charge is
\begin{eqnarray}
\label{eq:qdef}
  Q \equiv  {-i \over 2\pi} \sum_p \ln U_p.
\end{eqnarray}
The result for these quantities is known in terms of modified Bessel functions
for finite $\beta$ and $V$~\cite{Kovacs_1995,Bonati:2019ylr,Bonati:2019olo}:
\begin{align}
\begin{split}
P =& {\sum_n I'_n[\beta] I_n[\beta]^{V-1}\over \sum_n I_n[\beta]^V}, \\
\chi_t =&-{\sum_n A_n(\beta) I_n(\beta)^{V-1}\over \sum_n I_n(\beta)^V} \\&- (V-1) {\sum_n B^2_n(\beta) I_n(\beta)^{V-2}\over \sum_n I_n(\beta)^V
},
\label{eq:th}
\end{split}
\end{align}
where
\begin{eqnarray}
A_n(x) &\equiv & -{1 \over 2 \pi} \int_{-\pi}^\pi \left({\phi\over 2 \pi}\right)^2 e^{i n \phi+ x \cos \phi} d \phi,\\\nonumber
B_n(x) &\equiv & {i \over 2 \pi} \int_{-\pi}^\pi {\phi\over 2 \pi}\ e^{i n \phi+ x \cos \phi} d \phi,
\end{eqnarray}
and the sums in $n$ are over all integers. 
The infinite volume limits are
\begin{eqnarray}
\lim_{V\rightarrow\infty} P={I_1(\beta)\over I_0(\beta)}, ~~~\lim_{V\rightarrow\infty} \chi_t=-{A_0(\beta)\over I_0(\beta)}.
\end{eqnarray}
In the continuum limit,  $\beta\rightarrow \infty$, we recover the well-known results
\begin{align}
\begin{split}
\lim_{\beta\rightarrow \infty} \lim_{V\rightarrow\infty} P&=1-{\mathcal O}(\beta^{-1}), \\ \lim_{\beta\rightarrow \infty}\lim_{V\rightarrow\infty} \beta \chi_t&={1\over 4 \pi^2 } + {\mathcal O}(\beta^{-1}).
\end{split}
\end{align}

The partition function in fixed topology can also be easily derived from the
known partition function in the $\theta$ vacuum \cite{Kovacs_1995,Bonati:2019ylr,Bonati:2019olo}. At sufficiently large volume
\begin{eqnarray}
Z_Q(\beta,V) \equiv \int_{-\pi}^{\pi} d\theta e^{-i \theta Q}  Z_\theta(\beta, V),
\label{eq:zQ}
\end{eqnarray}
with \begin{eqnarray}
Z_\theta(\beta, V) = \left[I_{\theta\over 2\pi}(\beta)\right]^V. 
\label{eq:ztheta}
\end{eqnarray}
An interesting quantity is the average of the plaquette in fixed topology sectors, which shows a subtle dependence on $Q$, and it is analytically known also at finite $\beta$:
\begin{eqnarray}
 P_Q &\equiv& {1\over V} {1\over Z_Q} {d Z_Q\over d\beta} \nonumber\\
&=& {1\over Z_Q} \int_{-{1\over 2}}^{{1\over 2}} dz
 ~e^{-i 2 \pi z Q}   I'_{z}(\beta) \left[I_{z}(\beta)\right]^{V-1}.  \label{eq:PQ}
\end{eqnarray}
As we will see, this is a golden observable to test how algorithms perform in  sampling sectors of fixed-topology, given its high precision. 

\begin{figure*}[t]
	\centering
\includegraphics[width=12cm,keepaspectratio]{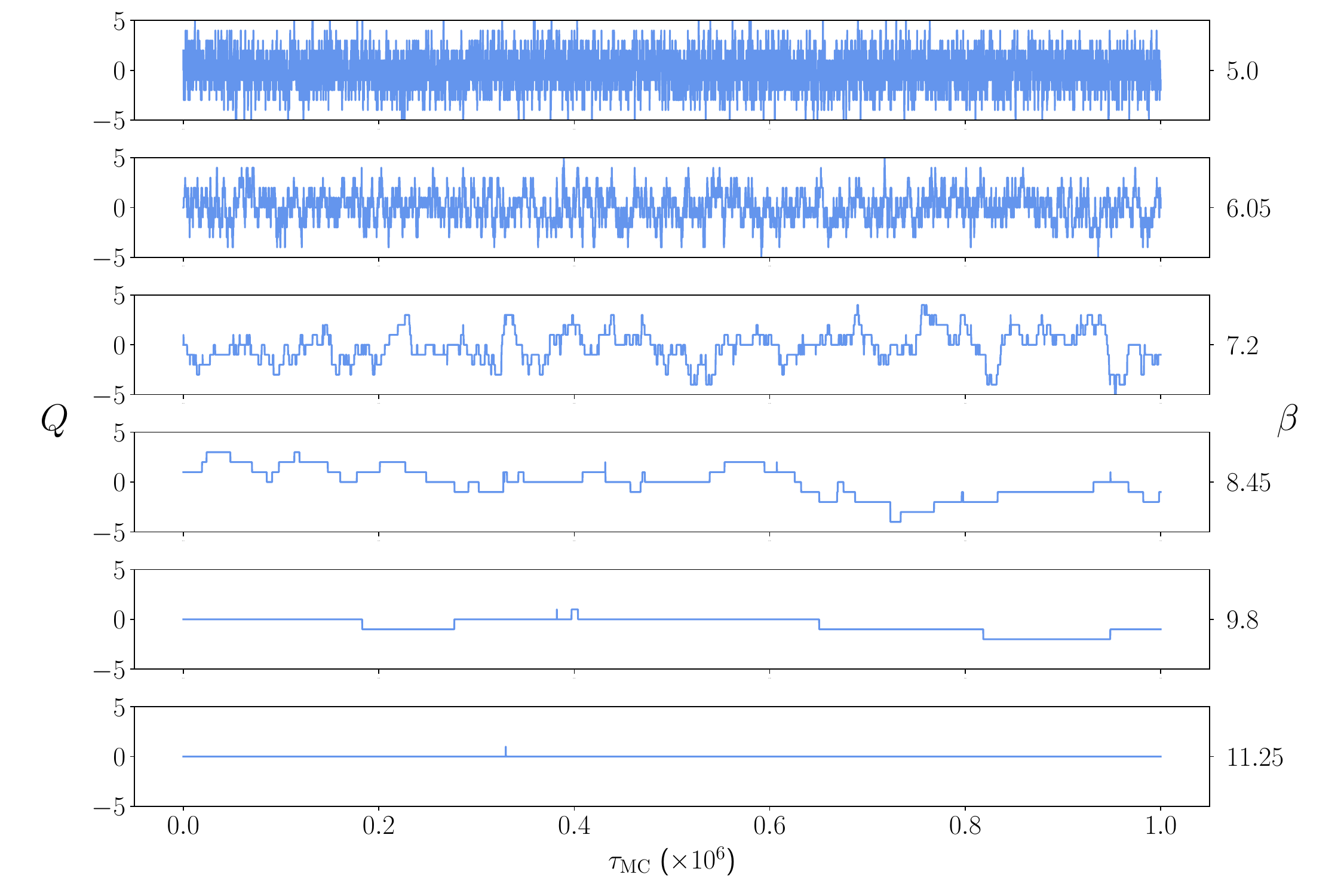}	
	\caption{Monte Carlo history of the topological charge $Q$ for increasing values of $\beta$ in a Markov chain of $10^6$ HMC configurations.}
	\label{fig:hmc_freeze}
\end{figure*}

\subsection{$N_f$ Schwinger Model}

In the theory with $N_f > 1$ fermions, the flavour symmetry group is $SU(N_f)_L\times SU(N_f)_R$. Even though spontaneous chiral symmetry breaking cannot occur in 2D and the condensate vanishes in the massless limit, the scaling of the condensate with the quark mass is non-trivial. 
The Gell--Mann--Oakes--Renner (GMOR) relation follows from the Ward identity (WI)
\begin{eqnarray}
F_{\pi}^2 M_\pi^2 = {2 m\Sigma(m)\over N_f},
\end{eqnarray}
where $m$ is the quark mass. The condensate is expected to scale with the quark mass \cite{Smilga:1992hx} as
\begin{eqnarray}
\Sigma(m) \propto m^{N_f-1 \over N_f+1} e^{2\over N_f+1},
\end{eqnarray} 
and therefore the pion mass scales with the quark mass as
\begin{eqnarray}
M_\pi^2 \propto m^{2 N_f\over N_f+1}.
\end{eqnarray}

The topological susceptibility vanishes in the limit of massless fermions. From the WI and the GMOR relation it follows
\begin{eqnarray}
\chi_t^{N_f} = {M_\pi^2 F_\pi^2\over 2 N_f}+{\mathcal O}(m^2),
\label{eq:xitNf}
\end{eqnarray}
and combining this with the Witten-Veneziano formula (and neglecting mass corrections to $F_\pi$), we expect
\begin{eqnarray}
\chi^{N_f}_t ={ {M_\pi^2 F_\pi^2\over 2 N_f}\over 1+  {M_\pi^2 F_\pi^2\over 2 N_f \chi_t^q}} ={1\over 4\pi\beta} { {M_\pi^2\beta}\over N_f+  \pi M_\pi^2\beta }, \label{eq:chit_interp}
\end{eqnarray}
which nicely interpolates between the pure gauge case, Eq.~(\ref{eq:xit}), for $M_\pi \rightarrow \infty$, and the flavoured result of Eq.~(\ref{eq:xitNf}), even though it is strictly derived close to the chiral limit.

\section{Winding HMC}

Even in this simple model, standard MC algorithms such as the HMC algorithm fail to reproduce the continuum limit expectations due to the bad sampling of topological sectors. In Fig.~\ref{fig:hmc_freeze} we plot an HMC history of the topological charge $Q$, showing the 
well-known topology freezing phenomenon. This results in the exponential growth of the autocorrelation time as $\beta \rightarrow \infty$ shown in Fig.~\ref{fig:tauQ2comp} for $Q^2$.

\begin{figure}[h!]
	\centering
	\includegraphics[width=0.75\linewidth]{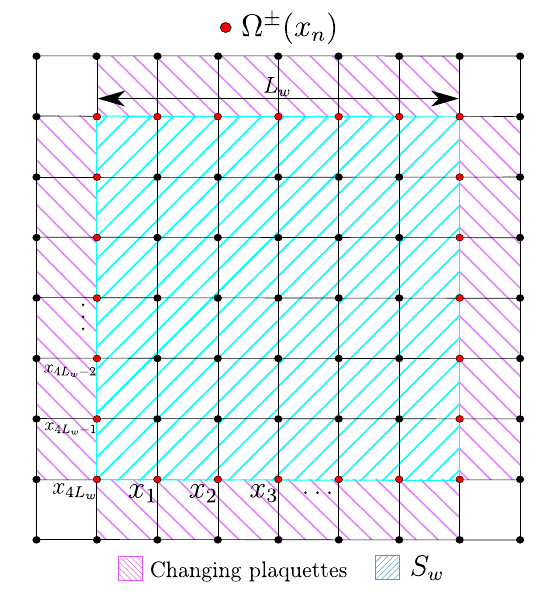}	
	\caption{Sketch of a winding transformation of size $L_w$. Gauge links whose starting and ending points are both in the blue region transform according to Eq.~(\ref{eq:windingtrafo}), while the others stay the same. Only the $1\times 1$ Wilson loops in the violet region change. } \label{fig:winddraw}
\end{figure}

The basic idea of our proposal is to combine HMC steps with a Metropolis--Hastings accept-reject step, where the trial configuration is obtained from the previous one by performing a winding. The winding transformation acts on the link variables whose starting and ending points are within a square region $S_w$ of size $L_w$,
\begin{align}
\begin{split}
U_\mu(x) \rightarrow U^\Omega_\mu(x) &\equiv \Omega(x) U_\mu(x) \Omega^\dagger(x+\hat\mu),   \label{eq:windingtrafo}
\end{split}
\end{align}
if both $x, x+\hat \mu \in S_w$. By contrast, the other links remain unchanged. The anatomy of the winding step is depicted in Fig.~\ref{fig:winddraw}.


We only have to define the gauge transformation $\Omega$. We set $\Omega = 1$ except at the boundary where it is chosen to have a winding number. If $x_n$ are the points on the boundary of $S_w$, ordered from $n=1,\dots,4 L_w$, we pick 
\begin{eqnarray}
\Omega^\pm(x_n) = e^{\pm i  {\pi \over 2}{n \over L_w}},
\end{eqnarray}
where the $+$ denotes a winding and the $-$ an antiwinding. The sign is chosen with 50\% probability, and is common for the $n$ points, ensuring that the transformation will yield a change in the topological charge of $\Delta Q = \pm 1$ in smooth configurations. The invariance of the measure ensures that this transformation has a trivial Jacobian, $dU^\Omega = dU$.  

The transition probability for $U\rightarrow U'$ of this Metropolis--Hastings step is
\begin{align}
&q(U'|U) = T(U\rightarrow U') p_{\rm acc}(U'|U) \nonumber\\
&+ \delta(U'-U) \sum_{U''} T(U\rightarrow U'') \left(1-p_{\rm acc}(U''|U)\right),
\label{eq:q}
\end{align}
with
\begin{eqnarray}
T(U \rightarrow U') = {1\over 2} \delta(U'- U^{\Omega^+}) + {1\over 2} \delta(U'-U^{\Omega^-}).
\label{eq:t}
\end{eqnarray}
Since $q(U'|U) = q(U|U')$ due to the 50\% probability of performing a winding or antiwinding, $p_{\text{acc}}$ is just the usual Metropolis \cite{Metropolis53} accept-reject probability
\begin{align}
\begin{split}
p_{\rm acc}(U'|U) =&\, {\rm min}\left\{ 1,{p(U')\over p(U)}\right\}, \\
{\rm with }\quad p&(U) = e^{-S[U]}
\end{split}
\end{align}
being the target probability distribution. In the pure gauge theory $S[U]$ is the plaquette action, whereas in the dynamical theory it includes the fermionic determinant. The latter is evaluated stochastically using one pseudofermion.

It is easy to check that $p(U)$ is the equilibrium distribution of such a Markov chain, i.e.,
\begin{align}
\int dU p(U) q(U'|U) &= p(U'), \label{eq:conv}\\ 
\int d U' q(U'|U) &= 1.
\end{align}
Substituting Eq.~(\ref{eq:q}) into Eq.~(\ref{eq:conv}), we get
\begin{align}
\begin{split}
	\int dU p(U) q(U'|U) =& {1\over 2} \sum_{\Omega=\Omega^\pm} \left[ \vphantom{\frac{1}{2}} p(U^{\prime  \Omega}) p_{\rm acc}(U'|U^{\prime  \Omega}) \right.
			    \\&\left.+ p(U') (1-p_{\rm acc}(U^{\prime  \Omega}|U')) \vphantom{\frac{1}{2}} \right] \\
=&\,p
(U'),
\end{split}
\end{align}
where the last step can be easily obtained after considering the two cases $p(U')<p(U^{\prime  \Omega})$, or $p(U')>p(U^{\prime  \Omega})$ for each $\Omega$. 

By itself this algorithm is obviously not ergodic, since only a
predefined change is performed. Ergodicity should be  
ensured by combining one or several winding steps with a standard HMC
step. We refer to this algorithm as wHMC. 

\subsection{Pure Gauge Case} \label{sec:puregauge}

 We have carried out a simulation of this algorithm for volumes with fixed $V/\beta \sim 80$
at various values of the lattice spacing, $\beta$, for the pure gauge theory.  Tab.~\ref{tab:comp} includes the parameters and results from these simulations for both HMC and wHMC. 
\begin{table*}[ht!]
    \centering
    \begin{tabular}{r|c|c|c|c|c|c|c|c}
        &{ $L_w$}   
        & \multicolumn{1}{c|}{$\beta$} & \multicolumn{1}{c|}{$L$}  &$\langle P \rangle$ & $\tau_P$ &  \# jumps& $\langle Q^2\rangle$ & $\tau_{Q^2}$ \\ \hline \hline
         HMC & --- & \multirow{2}{*}{5}& \multirow{2}{*}{20}  &  0.893439(26) &
         5.6(2) &  $1.9\cdot 10^4$ & 2.282(51) & 122(8) \\ 
         wHMC & 10 &  &   &  0.893410(35) & 5.2(2)  & $1.5\cdot 10^5$ & 2.359(19) & 7.6(2) \\ \hline
        \hline
         HMC &--- &  \multirow{2}{*}{6.05} & \multirow{2}{*}{22}  & 0.913132(24)
             & 9.4(2)  &$ 3 \cdot 10^3$ & 2.140(95) & 500(60) \\ 
         wHMC &11& &    &  0.913146(32) & 8.2(3) & $1.3\cdot 10^5$ & 2.219(18) & 7.9(2) \\ \hline
        \hline 
         HMC &--- &\multirow{2}{*}{7.2} & \multirow{2}{*}{24}  &  0.927675(20) &
         10.4(4)  & 434 & 2.53(35) & $5.0(14)\cdot 10^3$ \\ 
         wHMC &12&  &  &  0.927665(25) & 8.8(3)  & $1.2 \cdot 10^5$ & 2.201(19) & 8.8(2) \\ \hline\hline
        { HMC} & ---&\multirow{2}{*}{8.45}  & \multirow{2}{*}{26}   &
        0.938822(14) & 9.2(2)  & 56 & 2.21(78) & $40(20) \cdot 10^3$ \\ 
        { wHMC} & 13& &  & 0.938806(20) & 8.9(4)  & $1.1\cdot 10^5$ & 2.183(20) & 10.2(3) \\ \hline\hline
        { HMC} & --- & \multirow{2}{*}{9.8} & \multirow{2}{*}{28}  &
        0.947581(12) & 10.4(4)  & 9 & 0.84(39) & $80(40)\cdot 10^3$ \\ 
        { wHMC} &14& &   &  0.947539(17) & 10.7(5)  & $10^5$ & 2.141(20) & 10.6(3) \\ \hline\hline
        { HMC} & ---&\multirow{2}{*}{11.25} & \multirow{2}{*}{30} &  0.954519(11) & 7.0(3)  & 2 & --- & --- \\
       { wHMC} &15& &  &  0.954455(15) & 12.3(6)  & $0.9 \cdot 10^5$ & 2.168(21) & 11.9(3) \\  \hline \hline
    \end{tabular}
    \caption{Simulation parameters and results for the pure gauge model using $
    N_{\text{conf}}=5 \cdot 10^5$ configurations for wHMC and $10^6$ for HMC. The column ``\# jumps'' indicates the number of transitions in which the topological charge changes by at least one unit.  The integrator of the HMC step is tuned such that the acceptance is $\sim 90\%$. } 
    \label{tab:comp}
\end{table*}

\begin{figure}
	\centering
	\includegraphics[width=9cm,keepaspectratio]{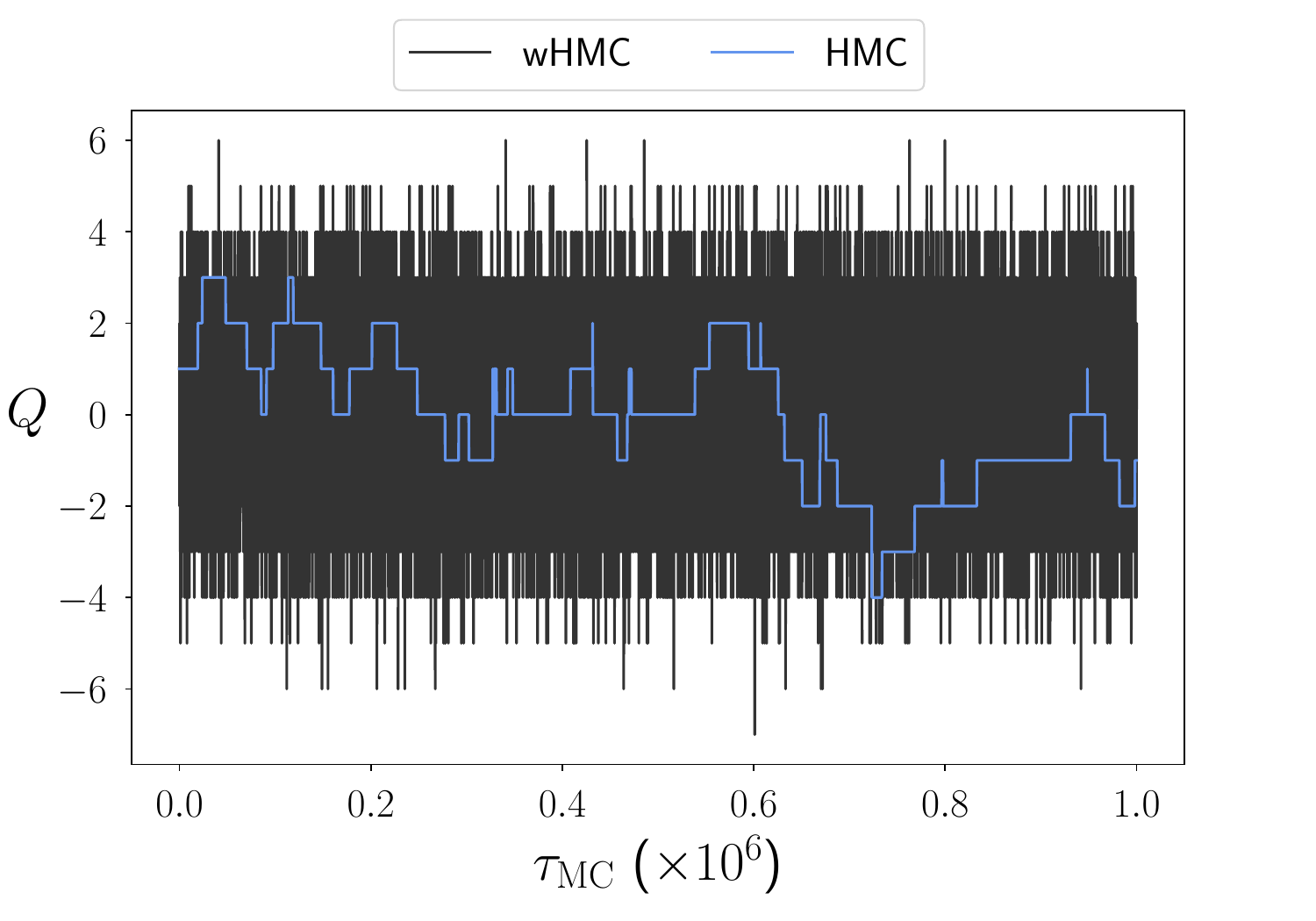}	
	\caption{Monte Carlo history of the topological charge for the two algorithms with $10^6$ configurations  at $\beta=8.45$.}
	\label{fig:compPQ2}
\end{figure}

\begin{figure}
	\centering
	\includegraphics[width=9cm,keepaspectratio,angle=0]{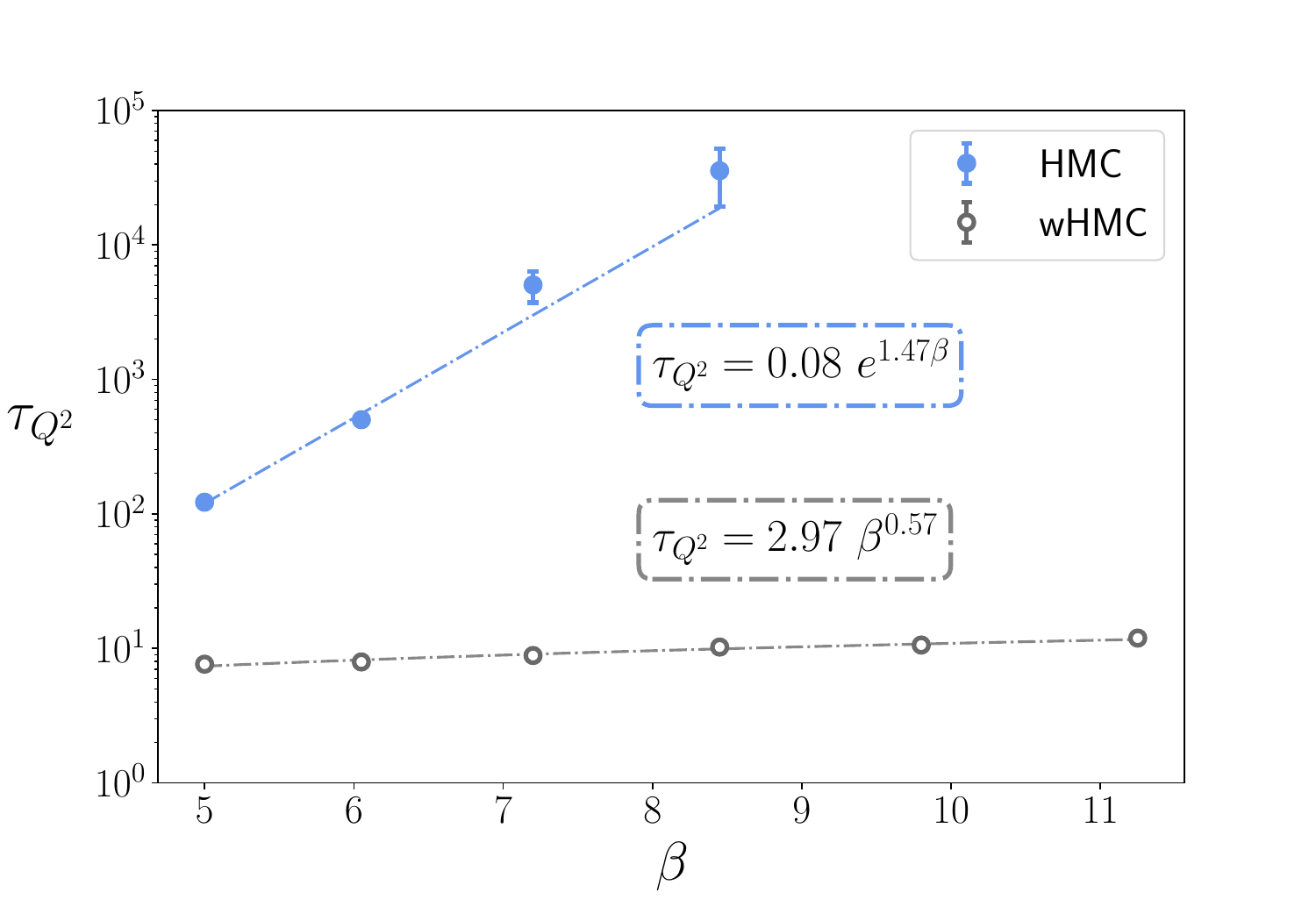}	
	\caption{Autocorrelation time of $Q^2$ as a function of $\beta \propto 
	a^{-2}$, as obtained with the wHMC and HMC algorithms.}
	\label{fig:tauQ2comp}
\end{figure}

Two MC histories of the topological charge for HMC and wHMC are compared in
Fig.~\ref{fig:compPQ2}, where the freezing of topology is absent for wHMC. This
can be quantified more precisely by looking at the scaling of the
autocorrelation times with $a^{2} \sim \beta^{-1}$. As shown
in Fig.~\ref{fig:tauQ2comp}, there is an enourmous improvement with respect to
standard HMC. The curves in Fig.~\ref{fig:tauQ2comp} correspond to the
two-parameter fits:
\begin{equation}
\tau_{Q^2}(\beta) \big|_{\rm HMC} = A \exp( b \beta),\quad \ \tau_{Q^2}(\beta) \big|_{\rm wHMC} = A \beta^{b}.
\end{equation}
The best fit parameters are $b=1.47(14)$ for the exponential fit to HMC, and
$b=0.565(53)$ for the power-law fit to wHMC. Therefore we find an exponential
scaling of the autocorrelation time with the lattice spacing for HMC, in
agreement with previous findings
\cite{Alles:1996vn,DelDebbio:2002xa,DelDebbio:2004xh,Schaefer:2010hu}, versus a
scaling proportional to $\sim \sqrt{\beta }$ for wHMC.

\begin{figure*}
	\centering
	\includegraphics[width=10.2cm,keepaspectratio,angle=0]{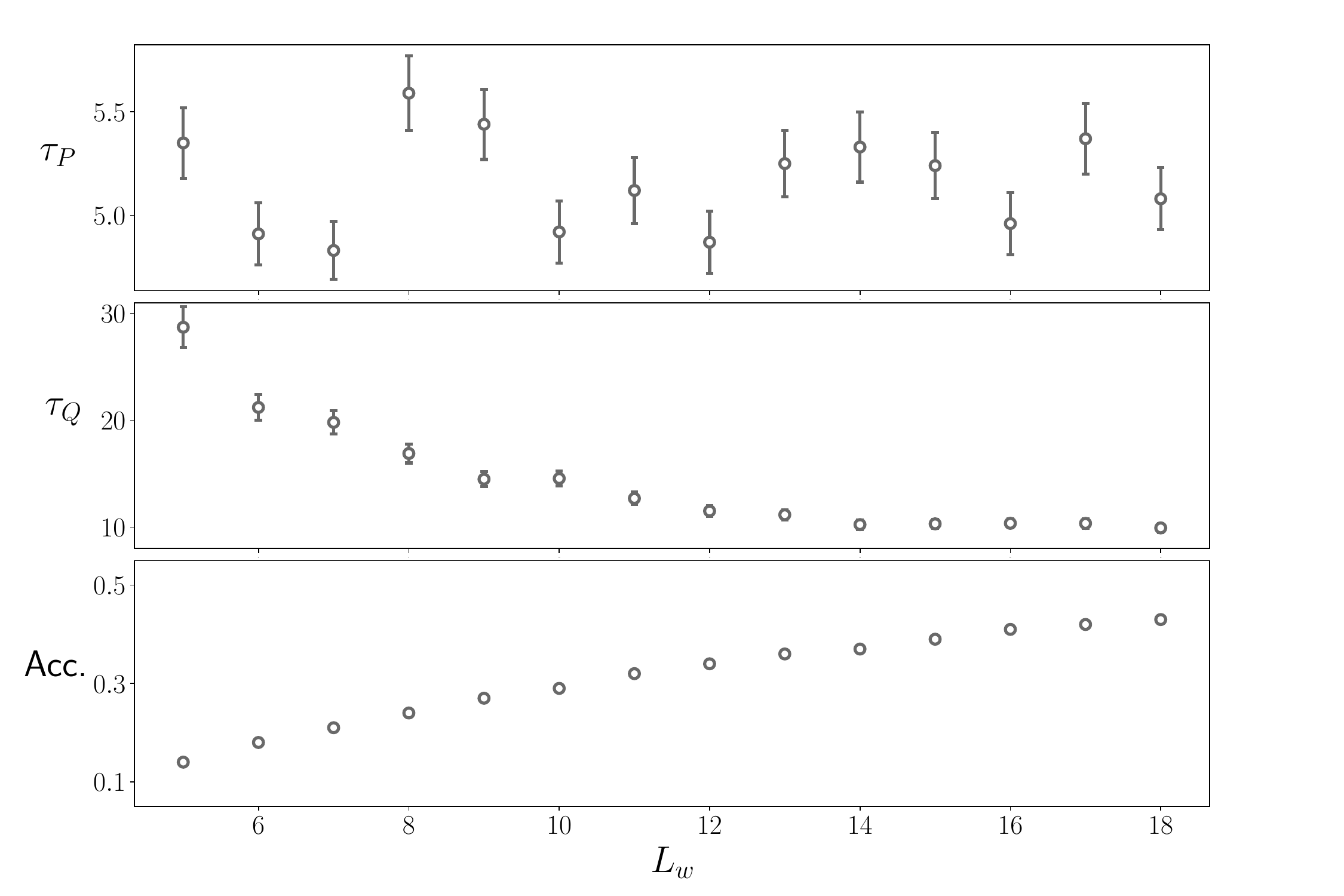}	
	\caption{Dependence on the winding size, $L_w$, of the plaquette autocorrelation time (top panel), the topological charge autocorrelation time (middle) and the acceptance of one winding step (bottom pannel),  at $\beta=5$. }
	\label{fig:Lwdep}
\end{figure*}

We have also studied the dependence on the size of the winding region, $L_{w}$. In Fig.~\ref{fig:Lwdep} we show the autocorrelation times for $P$, $Q$  and  the acceptance rate of the winding step as function of $L_{w}$: the acceptance of the winding grows with $L_{w}$ reaching $50\%$ at the largest $L_{w}$ considered, and $\tau_Q$ improves in a correlated fashion, while $\tau_P$ is insensitive to $L_{w}$. 


\begin{figure}
	\centering
	\includegraphics[width=\linewidth,keepaspectratio,angle=0]{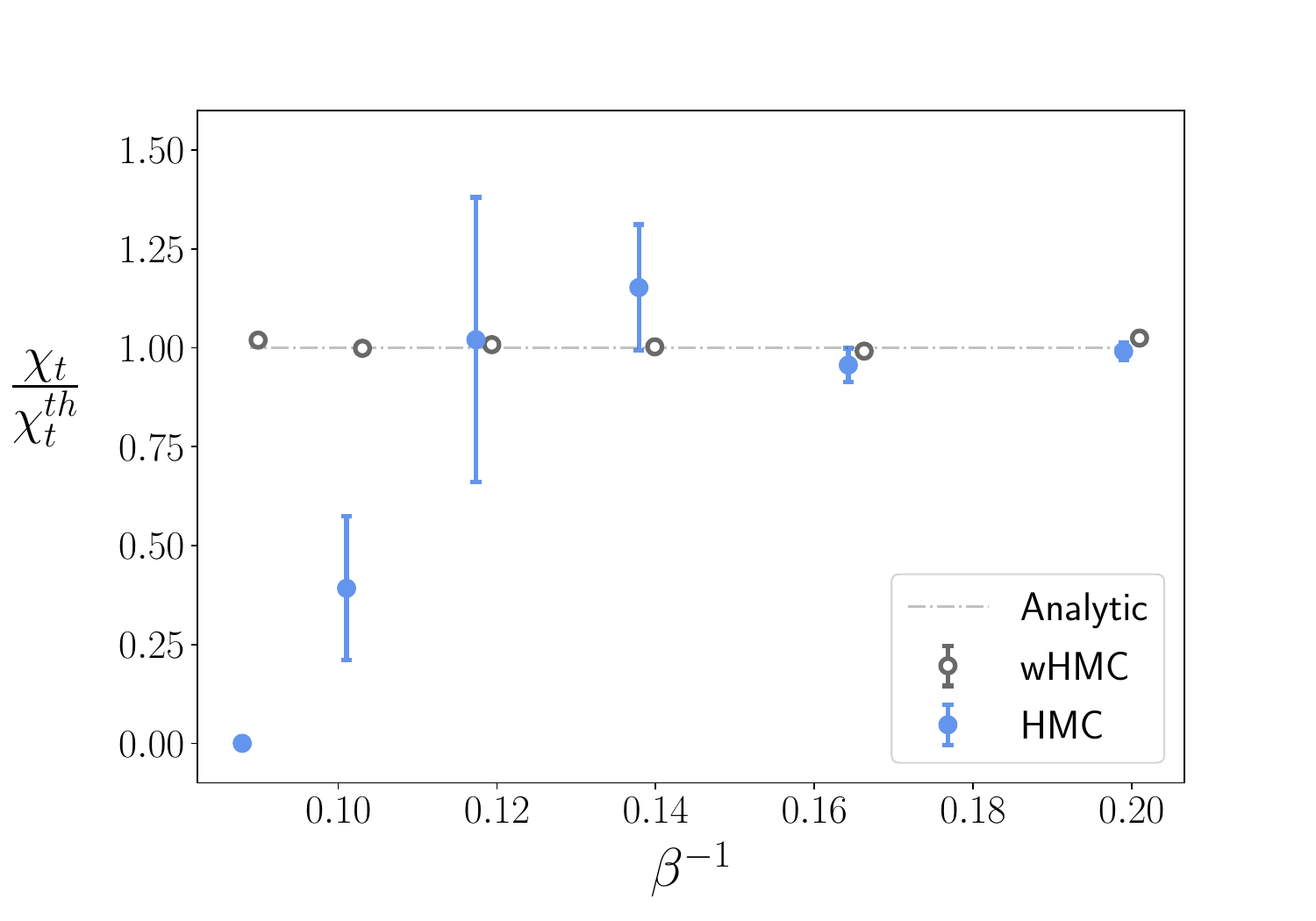}

	\includegraphics[width=\linewidth,keepaspectratio,angle=0]{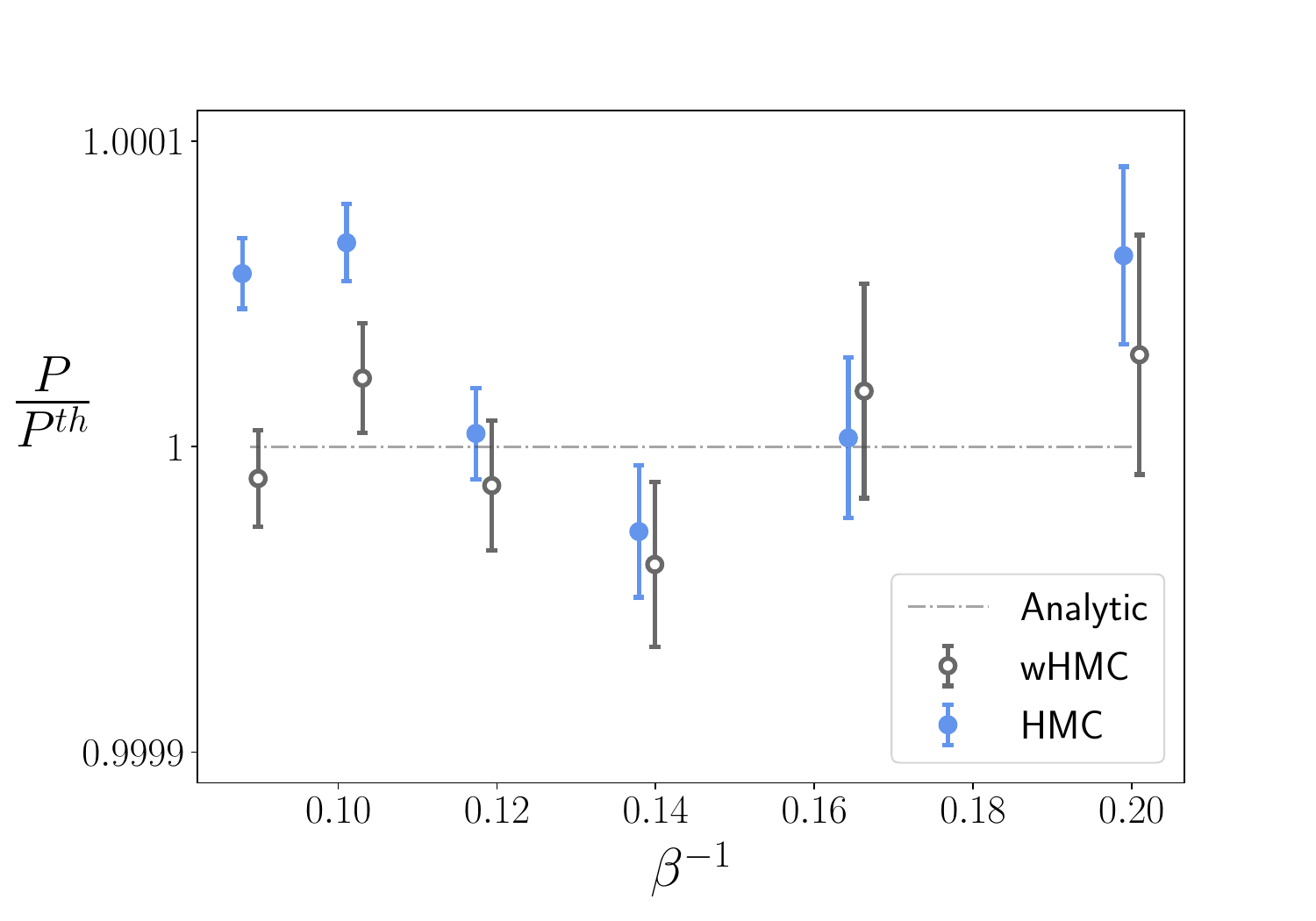}	
	\caption{ Average topological susceptibility (top) and plaquette (bottom) normalized to the exact result of Eqs.~(\ref{eq:th}) as a function of $\beta^{-1}$ for both algorithms.  }
	\label{fig:results}
\end{figure}

One can understand the improvement of the acceptance rate of the winding step
with $L_{w}$. The change in the action when a winding is performed is restricted
to the plaquettes at the boundary of $S_{w}$, and is due to the change in the
links at the boundary---see violet region in Fig.~\ref{fig:winddraw}. The change
in the phase of the plaquette, $\delta \theta_{p}$, due to the transformation
$\Omega^{\pm}$ is therefore $\mp \frac{\pi}{2L_{w}}$. We refer to $\delta S_{w}$
as the outer boundary of the winding region. For sufficiently large $L_{w}$, the
change in the phase of the plaquette is small and we can approximate
\begin{align}
\Delta S \approx  - \beta\frac{\pi}{2L_{w}} \sum_{p\subset \partial S_{w}}^{} \sin(\theta_{p})+ \beta \frac{\pi^2}{4L_{w}^2} \sum_{p \subset \partial S_{w}}^{}
\cos(\theta_{p}).
\end{align}
The average of the first term of $\Delta S$ vanishes, while the last term
averages to
\begin{align}
\langle \Delta S \rangle \simeq \frac{\beta \pi^2}{L_{w}}.
\end{align}
The acceptance increases as $L_{w} \rightarrow \infty$ at fixed $\beta$, since the change in the action averages to zero. We therefore conclude that the most efficient approach is to set the winding size to the largest possible value in this case.

The result for the average plaquette and the topological susceptibility normalized to 
the analytical results of Eqs.~(\ref{eq:th}) are shown in Fig.~\ref{fig:results}. The agreement with the exact results for both observables is very good for the wHMC algorithm, while for HMC both observables differ significantly from the theoretical expectation close to the continuuum limit. Although the divergence from the analytical result is more significant for the topological susceptibility, the plaquette also differs at various $\sigma$'s of confidence level.

\begin{figure}[h!]
	\centering
	\includegraphics[width=\linewidth,keepaspectratio,angle=0]{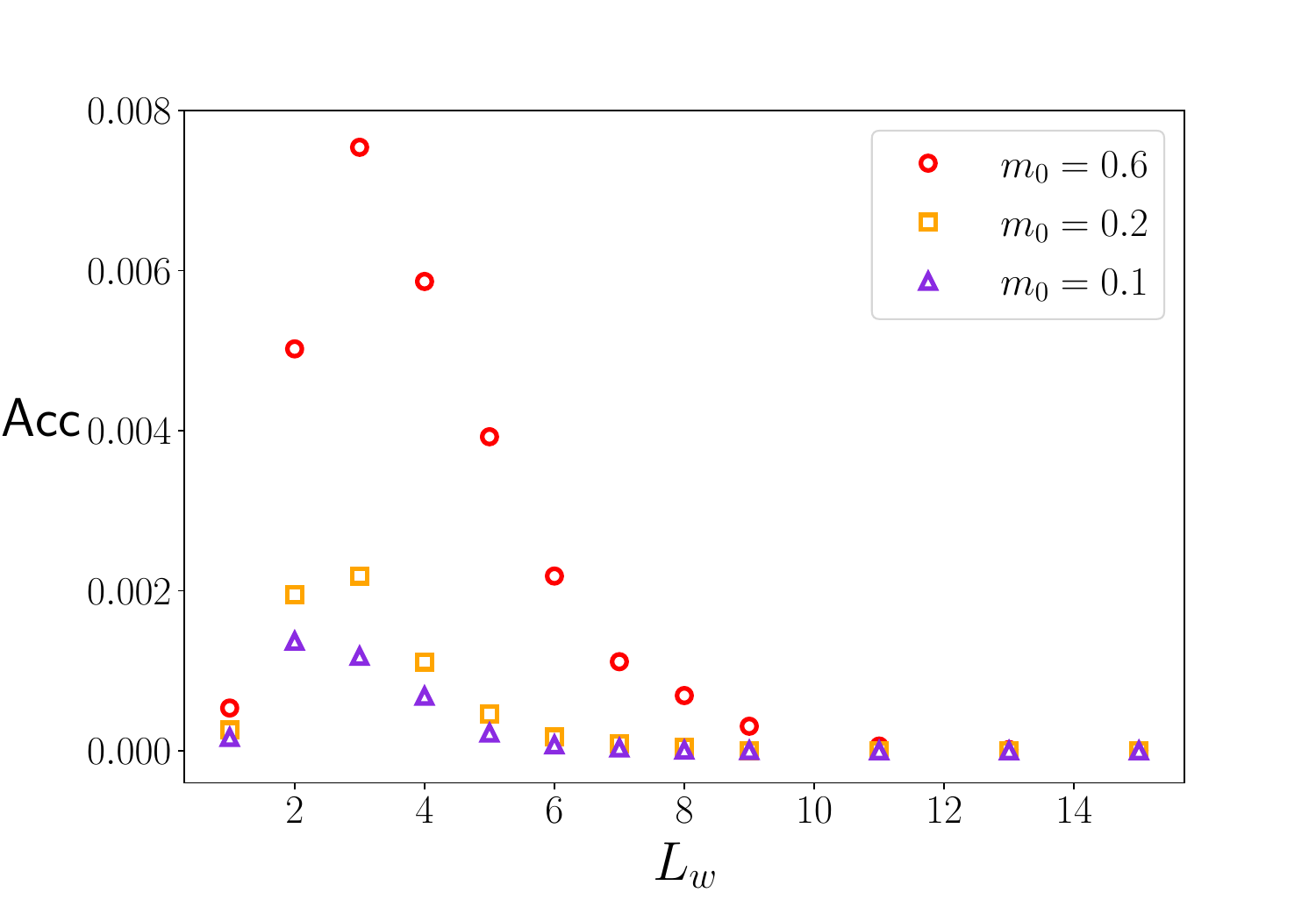}	
	\caption{ Acceptance of a winding step in wHMC as a function of $L_w$ for various bare quark masses at $\beta=5.0$. }\label{fig:AccvsLw}
\end{figure}

\subsection{$N_f=2$ Case}

\begin{figure}[h!]
	\centering
	\includegraphics[width=\linewidth,keepaspectratio,angle=0]{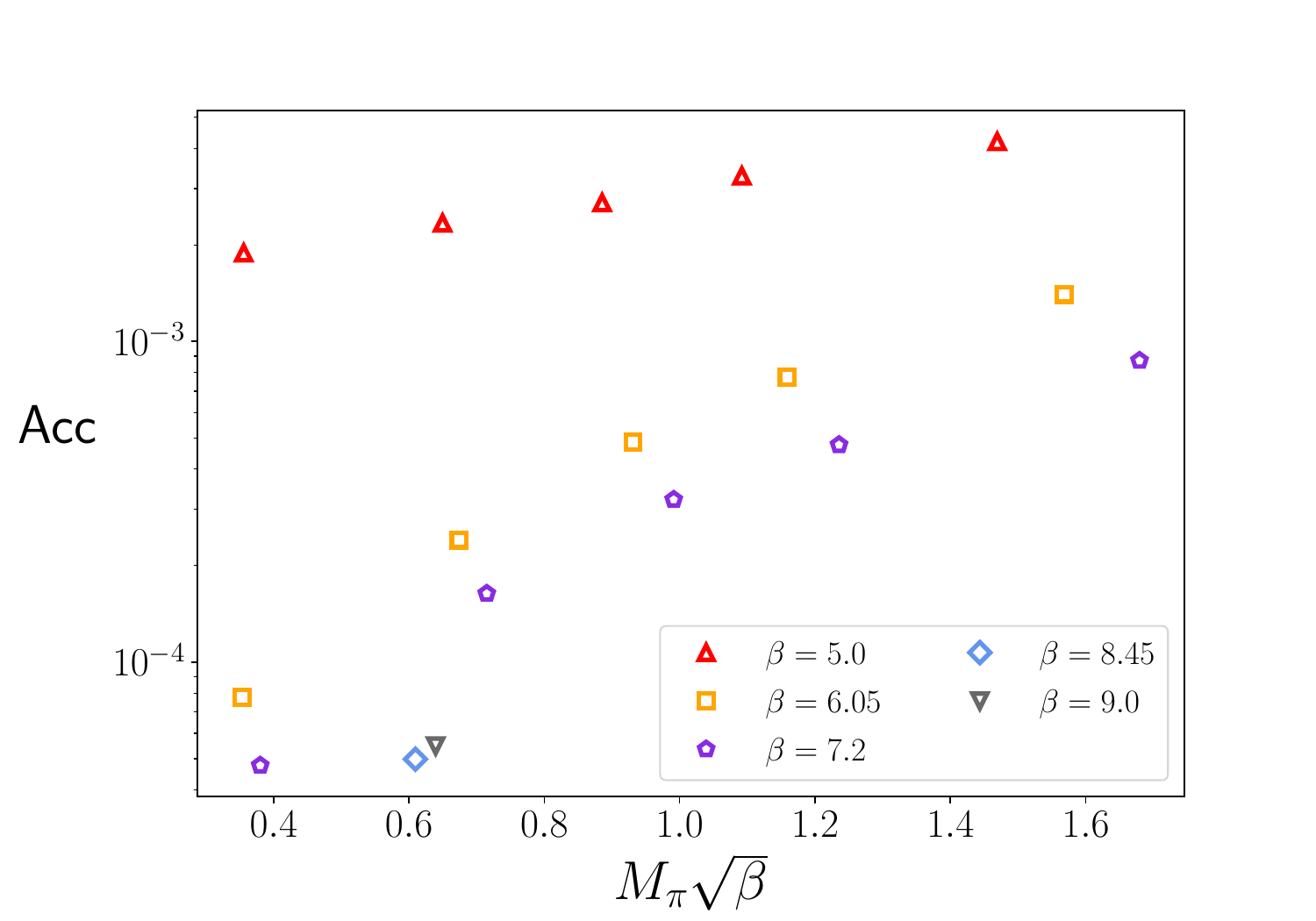}	
	\caption{  Acceptance of a winding step in wHMC as a function of $M_{\pi}\sqrt{\beta}$ for $L_w=3$ at various $\beta$'s.}
	\label{fig:Accvsm}
\end{figure}

The inclusion of the fermion determinant  is challenging in the wHMC algorithm, because the acceptance becomes very small. The reason is that  the change in the action induced by the winding can no longer be circumscribed to the boundary plaquettes, since the determinant is non-local. 
In Fig.~\ref{fig:AccvsLw} we show the acceptance of a winding Metropolis--Hastings step as a function of $L_w$.  The acceptance is seen to be below 1\%, and the highest acceptance is no longer reached for large $L_w$: the optimal $L_w$ is roughly 2-3 with a mild dependence on the quark mass. 
The value of the optimal acceptance is however very sensitive to $\beta$ and to the quark mass, decreasing as the chiral limit is approached. This is shown in Fig.~\ref{fig:Accvsm}, where we plot the acceptance as a function of the pion mass for fixed $L_w=3$ for various $\beta$.  There is indeed room for optimizing $L_w$ as a function of $\beta$ and the quark mass.

On the other hand, one winding accept-reject step involves one inversion of the Dirac operator, while an HMC step involves as many as the number of steps in the integrator, $n_{\rm HMC}$, which is of order ${\mathcal O}(10-100)$. Therefore, instead of one winding, we could perform $\mathcal{O}(100)$ winding accept-reject steps between HMC steps at a similar cost. A step of the wHMC is then defined as one HMC step + $n_{\rm W}$-winding accept-reject steps. This increases the computational cost of each wHMC step compared to a HMC one by a factor
\begin{equation}
r_{c} \equiv \frac{n_{\rm HMC} + n_{\rm W}}{n_{\rm HMC}}, \label{eq:cost}
\end{equation} 
while significantly improving the scaling with $\beta$ of the autocorrelation time of the topological charge.

We have performed a series of simulations at  several  $\beta$'s, computing the pion mass and the topological susceptibility for various values of the bare quark mass, $m_0$. The summary of our results is in Tab.~\ref{tab:conf_ferm}. 

\begin{table*}
    \centering
    \begin{tabular}{r|c|c|c|c|c|c|c|c|c}
             & $L$ & Configs. & $\beta$ & $m_0$ & $M_\pi$ & $\langle P \rangle$ & $\tau_P$ & $\langle Q^2 \rangle$ & $\tau_{Q^2}$ \\ \hline\hline
       wHMC  & 90 & 30000 & 5.0 & -0.05 & 0.15916(54)  & 0.899605(18) & 1.85(13) &  13.5(16) & 84(29) \\ \hline
        HMC   & 50 & 25000 & 5.0 & 0.0 & 0.29062(53)    & 0.898380(33) & 1.63(12)  & 6.59(68) & 71(25) \\ \hline
        wHMC  & 90 & 30000 & 5.0 & 0.0 & 0.29056(50)   & 0.898457(16) & 1.50(10) &  23.1(19) & 57(17) \\ \hline
        wHMC  & 90 & 30000 & 5.0 & 0.05 & 0.39600(50)  & 0.897577(17) & 1.63(11) &  28.0(30) & 81(27) \\ \hline
        wHMC  & 90 & 30000 & 5.0 & 0.1 & 0.48833(53)  & 0.896899(18) & 1.71(12) &  35.1(34) & 71(23) \\ \hline
        wHMC  & 90 & 30000 & 5.0 & 0.2 & 0.65709(49)  & 0.895959(17) & 1.66(11) &40.7(32) & 45(12) \\ \hline \hline
        wHMC  & 90 & 25000 & 6.05 & -0.04 & 0.14383(14)  & 0.917355(17) & 2.14(17) &  11.7(32) & 547(362 \\ \hline
        HMC   & 50 & 25000 & 6.05 & 0.01 & 0.27454(50)  &   0.916546(29) & 1.81(14) & 7.0 (17) & 522(343) \\ \hline
        wHMC  & 90 & 35000 & 6.05 & 0.01 & 0.27407(27)   & 0.916594(15) & 2.19(15) &  17.2(22) & 157(65) \\ \hline
        wHMC  & 90 & 35000 & 6.05 & 0.06 & 0.37853(38)   & 0.915966(14) & 1.97(13) &  30.3(26) & 87(28) \\ \hline
        wHMC  & 90 & 55000 & 6.05 & 0.11 & 0.47119(47)   & 0.915505(12) & 2.20(13)  & 32.6(39) & 172(62) \\ \hline
        wHMC  & 90 & 35000 & 6.05 & 0.21 & 0.63760(64)  & 0.914848(16) & 2.36(17)  & 33.9(24) & 38.5(92) \\ \hline \hline
        wHMC  & 90 & 53500 & 7.2 & -0.03 & 0.14174(36)   & 0.930709(10) & 2.34(14)  & 11.8(25) & 668(385) \\ \hline
        HMC   & 90 & 25000 & 7.2 & 0.02 & 0.26747(48)  &   0.930113(15) & 2.34(20) &  14.0(43) & 1026(345) \\ \hline
        wHMC  & 90 & 79450 & 7.2 & 0.02 & 0.26648(28) & 0.9301123(82) & 2.16(11) &16.5(16) & 261(95) \\ \hline
        wHMC  & 90 & 75000 & 7.2 & 0.07 & 0.36942(28)   & 0.9296760(88) & 2.32(12) &  21.6(19) & 178(58) \\ \hline
        wHMC  & 90 & 75000 & 7.2 & 0.12 & 0.46047(27)   & 0.9293559(87) & 2.24(11)  & 28.2(35) & 340(140) \\ \hline
        wHMC  & 90 & 95000 & 7.2 & 0.22 & 0.62595(23)   & 0.9289031(82) & 2.53(12) & 26.3(19) & 134(35) \\ \hline \hline
        HMC   & 50 & 200000 & 8.45 & 0.00 & 0.19743(17)    & 0.9407942(80) & 2.27(7) & 3.6(11) & 8820(3662) \\ \hline
        wHMC  & 50 & 35000 & 8.45 & 0.00 & 0.19720(41)  & 0.940812(18) & 2.02(14)  & 3.68(37) & 92(31) \\ \hline \hline
        HMC   & 50 & 85000 & 9.0 & 0.01 & 0.20972(20)    & 0.944408(11) & 2.07(1) &  -    & -  \\ \hline
        wHMC  & 50 & 25000 & 9.0 & 0.01 & 0.21362(55)   & 0.944381(20) & 2.10(17)  & 5.13(66) & 134(58) \\ \hline \hline
    \end{tabular}
        \caption{Simulation parameters and results from the $N_f=2$ simulations\label{tab:conf_ferm}.}
\end{table*}

\begin{figure}
	\centering
	\includegraphics[width=\linewidth,keepaspectratio,angle=0]{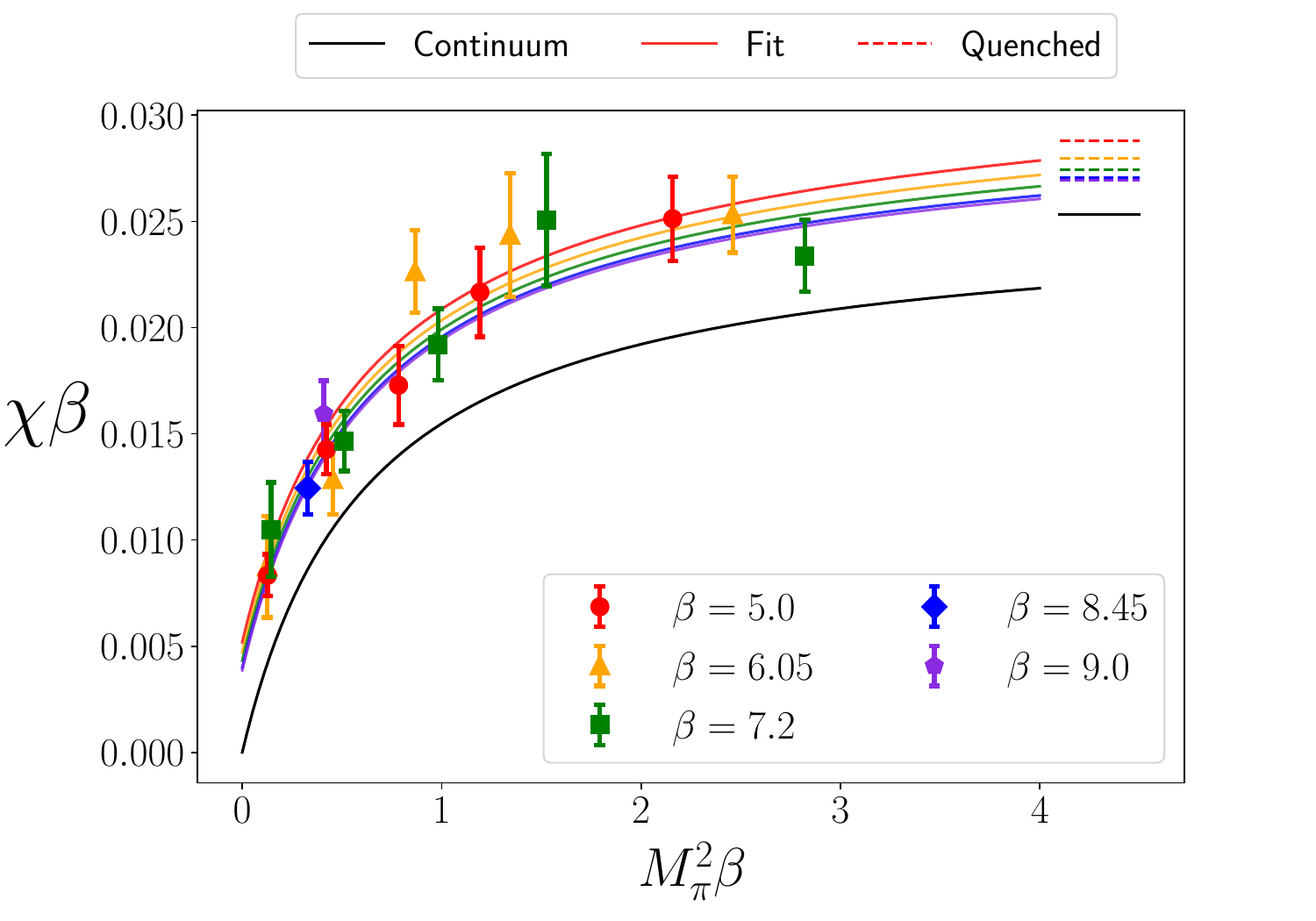}	
	\caption{  Topological susceptibility in the $N_f=2$ theory as a function of the pion mass. The coloured solid lines are fits to the expression in Eq.~(\ref{eq:fitchit}) for five $\beta$'s, while the black line is the continuum result. The horizontal dashed lines are the quenched expectations at the various $\beta$ and the continuum.}
	\label{fig:chiNfvsmpi}
\end{figure}

In Fig.~\ref{fig:chiNfvsmpi} we show the topological susceptibility as a function of the pion mass, together with the fit to the continuum expectation, Eq.~(\ref{eq:chit_interp}), plus generic cutoff effects that in the theory with unimproved Wilson fermions are expected to scale with ${\mathcal O}(a) \sim \beta^{-1/2}$. We fit the various results at various $\beta$ and quark masses to the ansatz
\begin{eqnarray}
\chi_t^{N_f=2} = {\rm Eq.(\ref{eq:chit_interp})} + (c + d  M_\pi^2) \beta^{-1/2} ,
\label{eq:fitchit}
\end{eqnarray}
where $c$ and $d$ are the fitting parameters. The agreement  of wHMC with the expectation is good, even at values of $\beta$ where the topology in HMC is completely frozen and does not allow to measure the topological susceptibility. Cutoff effects are significant and larger than in the pure gauge theory, as expected from the presence of Wilson fermions.

\begin{figure}
	\centering
	\includegraphics[width=\linewidth,keepaspectratio,angle=0]{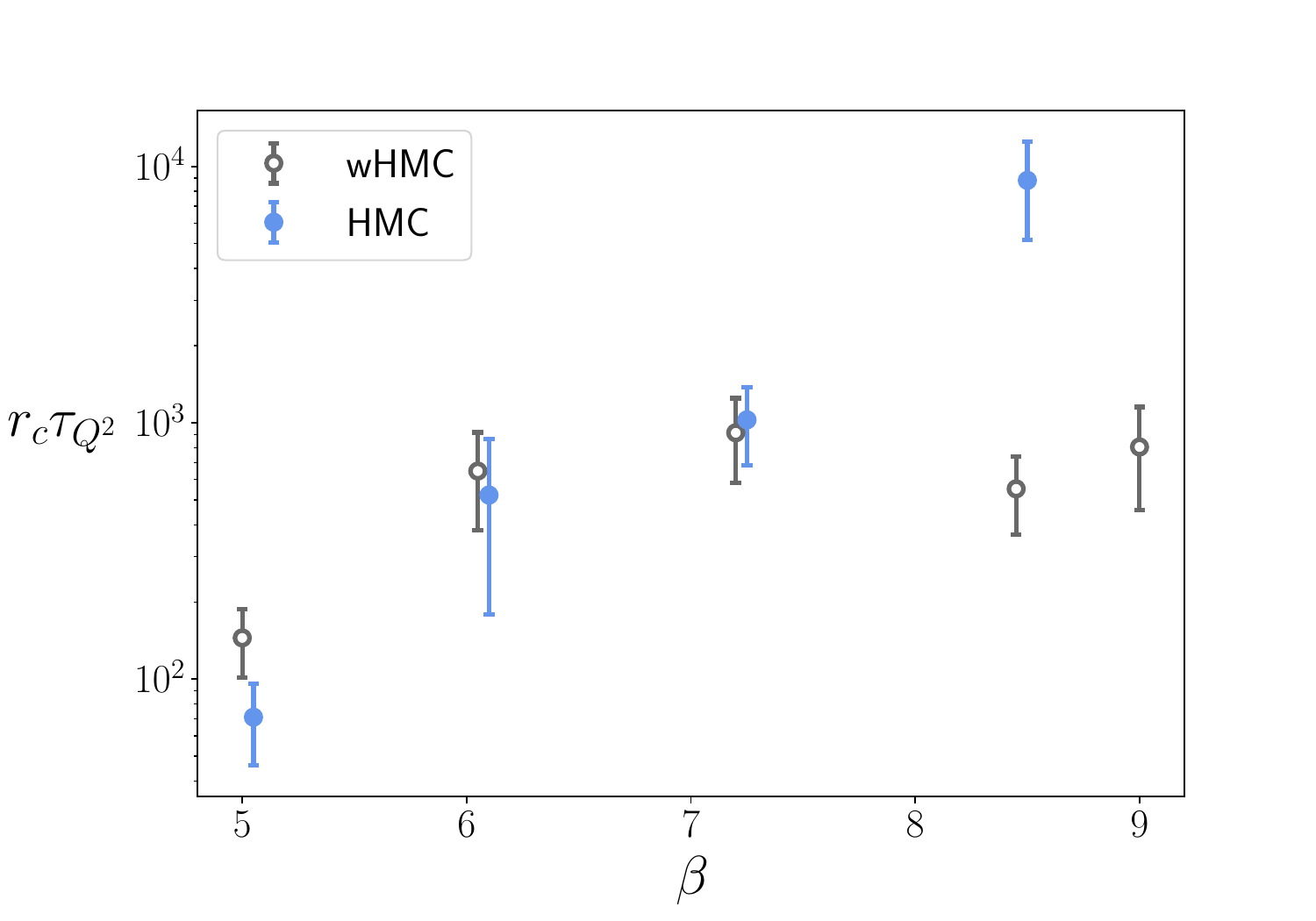}	
	\caption{  Scaling of the autocorrelation time for $Q^2$ times the cost ratio, Eq.~(\ref{eq:cost}), with $\beta$ for HMC and wHMC. The pion mass is kept approximately constant, $M_\pi \sqrt{\beta} \sim  0.65$. }
	\label{fig:ftauint}
\end{figure}

Even though the autocorrelation is larger than in the pure gauge theory, we still see a major improvement in the scaling towards the continuum limit as shown in Fig.~\ref{fig:ftauint}. Note that the autocorrelation time is multiplied by the factor in Eq.~(\ref{eq:cost}), which accounts for the increase in computational cost. In our simulations, this factor goes from $r_{c} \approx 2.53$ to $r_{c}=6$ in the range $\beta \in[ 5, 9 ]$.

\section{Results at fixed topological sector}
\label{sec:fixing-topology}

A way to overcome the topology freezing problem consists in extracting 
physical quantities of interest from simulations at fixed
topology and correcting for the finite-size dependence~\cite{Brower:2003yx,Aoki:2007ka} (see
also~\cite{Fritzsch:2013yxa} for applications in the context of finite
size scaling). A key ingredient in this
approach is that algorithms that suffer from topology freezing can
nevertheless sample correctly sectors of fixed topology. 
In this sense only the relative weights of different topological
sectors are difficult to compute for an algorithm suffering topology freezing. 
This hypothesis can be studied very accurately in the context of our
simple 2D model: on the one hand we can compare with the analytical results
in the pure gauge case, and on the other hand we can compare the
results with our wHMC algorithm for the $N_{f}=2$ case. 

\subsection{Pure Gauge}

\begin{figure}[h!]
	\centering
	\includegraphics[width=\linewidth,keepaspectratio]{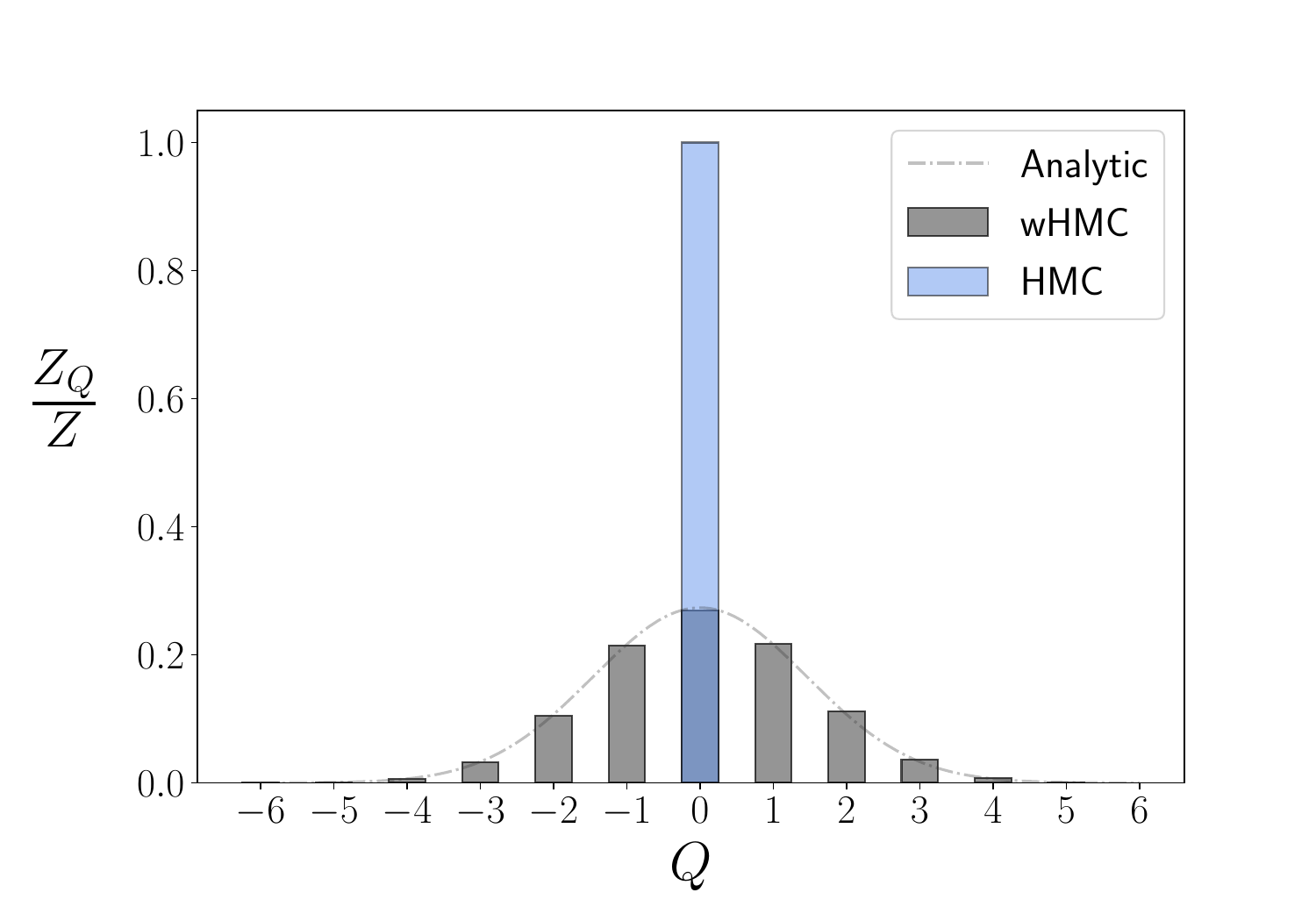}	
	\caption{ $Z_Q/Z$ versus $Q$ at $\beta=11.25$ for HMC and wHMC, versus the analytical result. }
	\label{fig:wq}
\end{figure}

Let us start with the pure gauge model. In
Fig.~\ref{fig:wq} we show the result for the weights of the different
topological sectors obtained with the two algorithms at $\beta=11.25$,
compared to the expectations in Eq.~(\ref{eq:zQ}). Clearly HMC fails
at evaluating these weights, while wHMC succeeds.   

 \begin{figure}[h!]
	\centering
	\includegraphics[width=\linewidth,keepaspectratio]{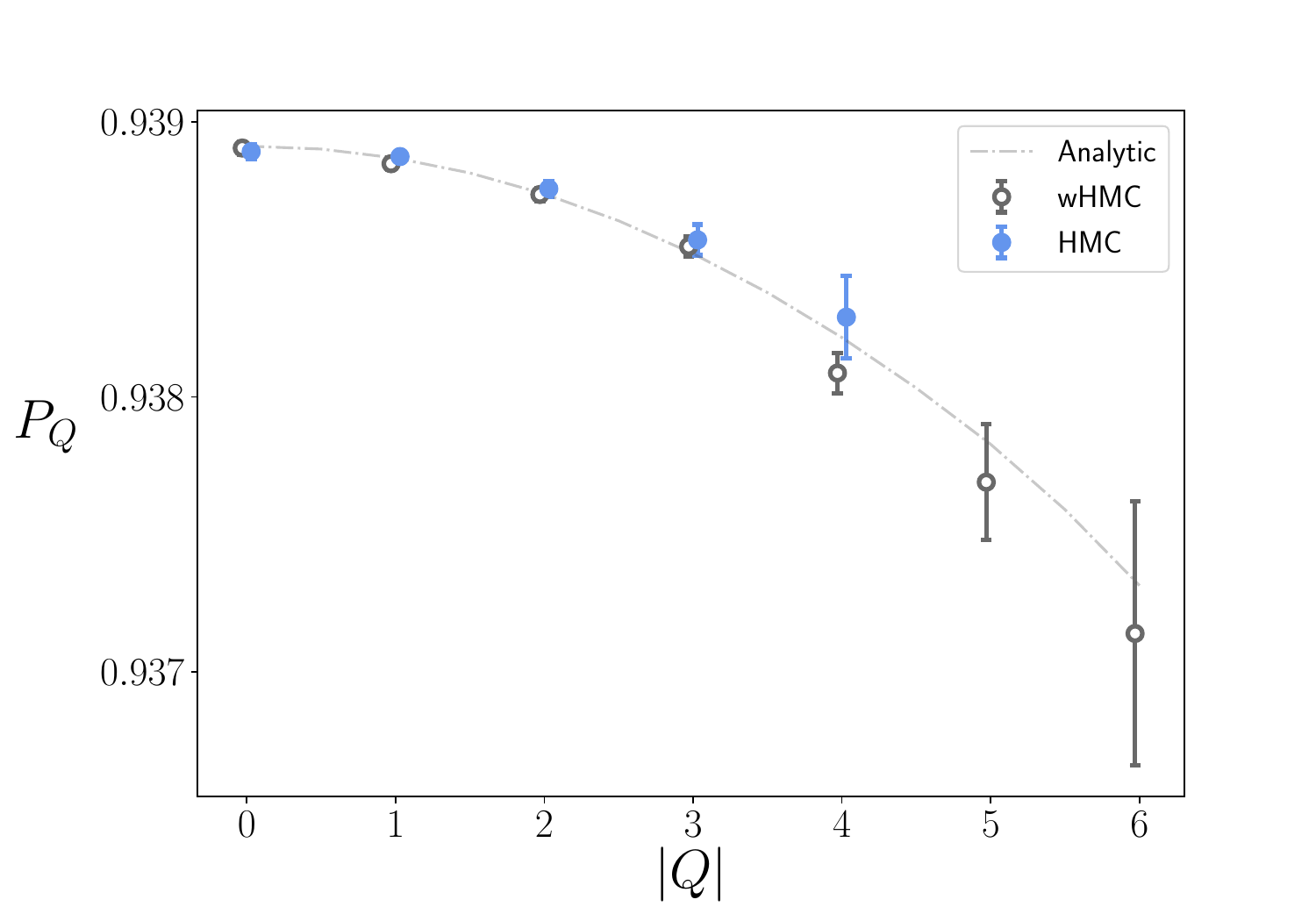}	
	\caption{ $ P_Q$ versus $Q$ at $\beta=8.45$ for wHMC and HMC versus the analytical result. }
	\label{fig:pq}
\end{figure}

\begin{figure}[h!]
	\centering
	\includegraphics[width=\linewidth,keepaspectratio]{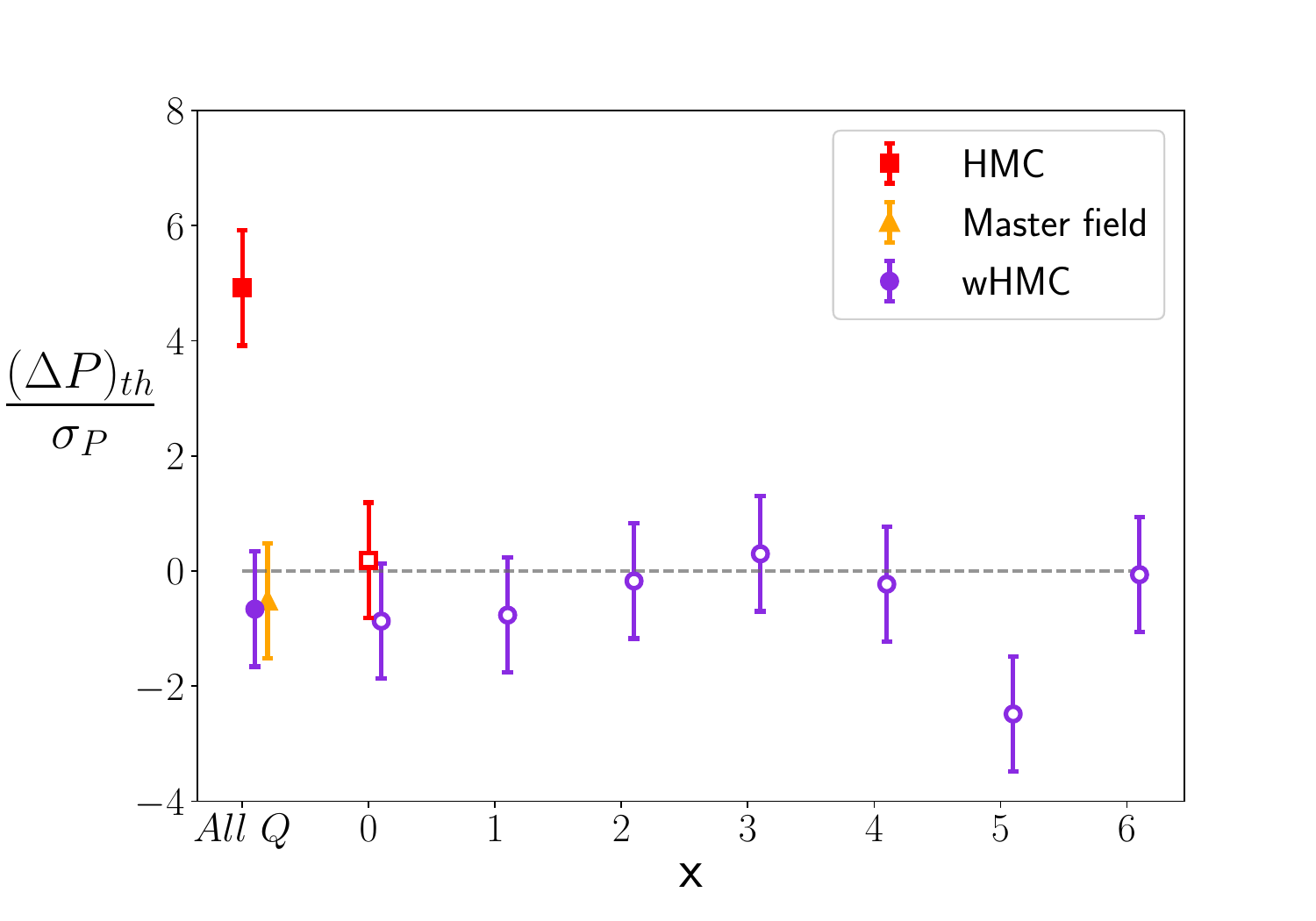}	
	\caption{ $ (\Delta P)_{\rm th} \equiv  P_{\rm x} - P^{\rm th}_{\rm x}$
	normalized with the standard deviation $\sigma$ for ${\rm x} = All\ Q$
(full symbols) or at fixed topology ${\rm x} = |Q|$ (open symbols), at $\beta=11.25$ for wHMC, HMC and master field, compared to the analytical result (dashed line). }
	\label{fig:pqdiff}
\end{figure}

In the pure gauge model, the plaquette at fixed topology has a small
but measurable $Q$ dependence (see Eq.~(\ref{eq:PQ}) and
Fig.~\ref{fig:pq}). We can therefore 
test whether the algorithm samples properly within each topological
sector and  reproduces the correct $Q$ dependence. We consider
the projected observable $O$ to the topological sector $n$  
\begin{equation}
O_n = \frac{ \langle  O \, \delta_{n}(Q)\rangle }{ \langle  \delta_{n}(Q)\rangle }, \label{eq:defPq}
\end{equation}
where
\begin{equation}
  \delta_{n}(Q) = \left\{
    \begin{array}{ll}
      1& |Q| = n\\
      0& \text{otherwise}
    \end{array}
    \right. .
\end{equation}

Fig.~\ref{fig:pqdiff} shows the difference between the measured
plaquette and the analytical expectation, $(\Delta P)_{\rm th} = P-P_{\rm
  th}$, in units of the error of the measured plaquette. 
We see that HMC fails to reproduce the correct expectation value
of the plaquette (label ``All $Q$'') by 5 standard deviations. 
This is expected since HMC is completely frozen and only $Q=0$ is
present in the Monte Carlo history, while the plaquette shows a small
(but noticeable) dependence on $Q$. 
On the other hand the expectation value of the plaquette projected to
the $Q=0$ sector is perfectly predicted by HMC. 
We also see that the wHMC, which is able to sample all topological
sectors, reproduces correctly the value of the plaquette projected to
all values of the charge from 0 to 6. 
It also reproduces the expectation value of the plaquette.

\subsection{$N_f=2$ results}

We now turn to the model with dynamical fermions. Simulations at fixed
topology have been performed in previous works using the HMC algorithm
for this
model~\cite{Czaban:2013haa,Czaban:2014gva,Bietenholz:2016ymo}.  

\begin{figure}[h!]
	\centering
	\includegraphics[width=\linewidth,keepaspectratio]{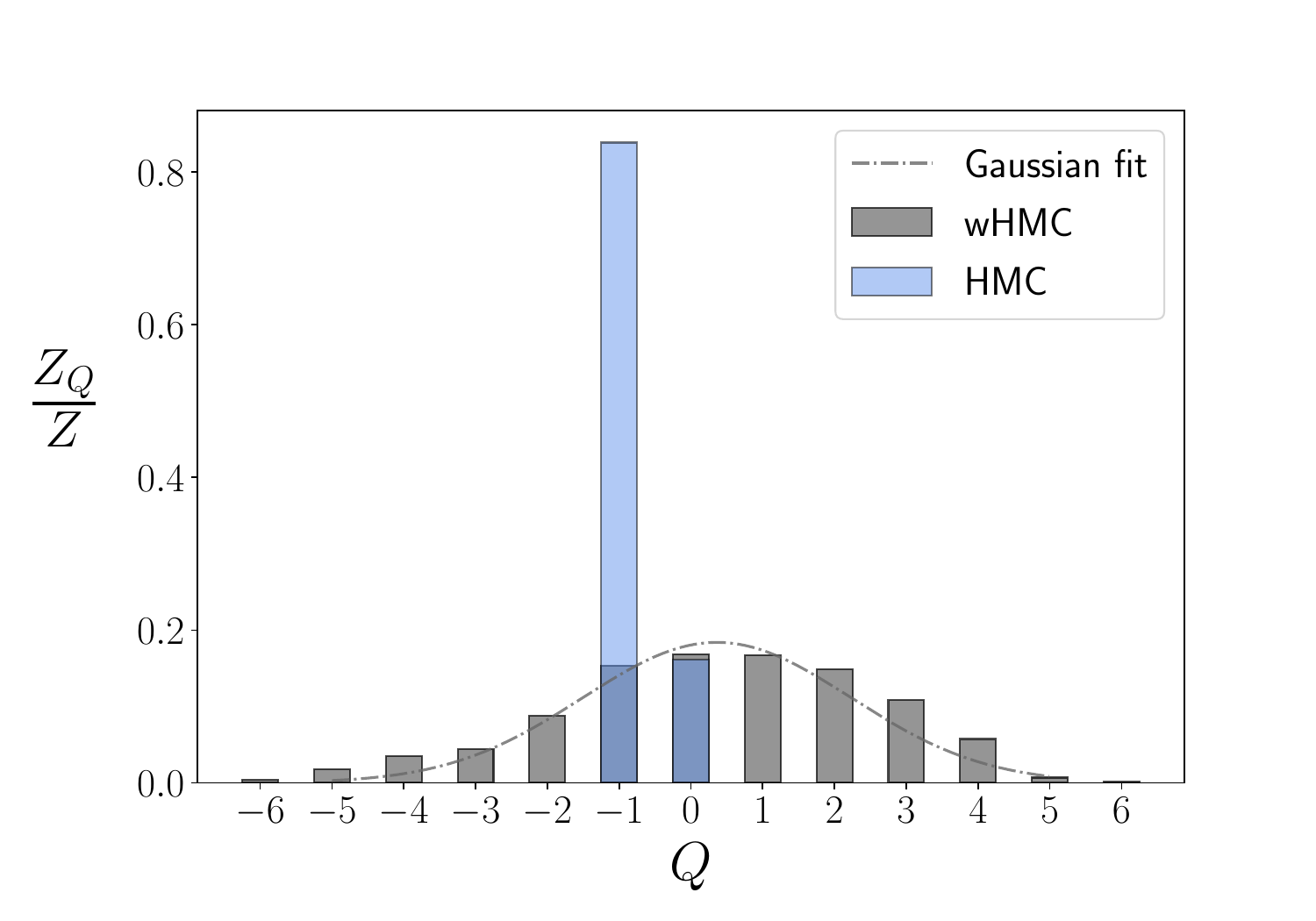}	
	\caption{ $Z_Q/Z$ versus $Q$ at $\beta=9.0$ for HMC and wHMC with $N_{f}
	= 2$. A Gaussian fit to the wHMC distribution is also shown. }
	\label{fig:wqferm}
\end{figure}

\begin{figure}[h!]
	\centering
	\includegraphics[width=\linewidth,keepaspectratio]{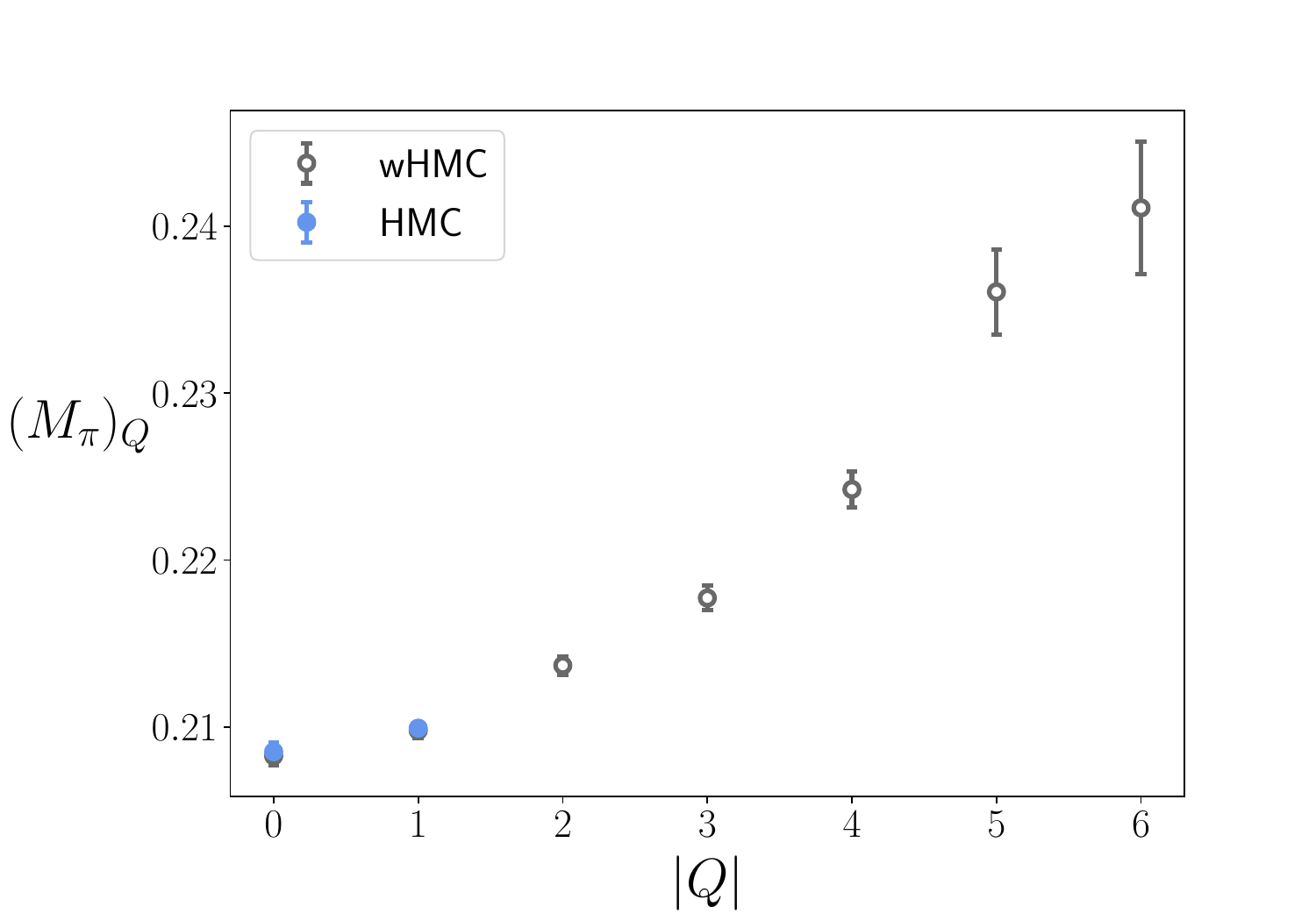}	
	\caption{ Pion mass at fixed topology, $(M_\pi)_Q$,  versus $Q$ at $\beta=9.0$ for wHMC and HMC. }
	\label{fig:MpiQ}
\end{figure}

\begin{figure}[h!]
	\centering
	\includegraphics[width=\linewidth,keepaspectratio]{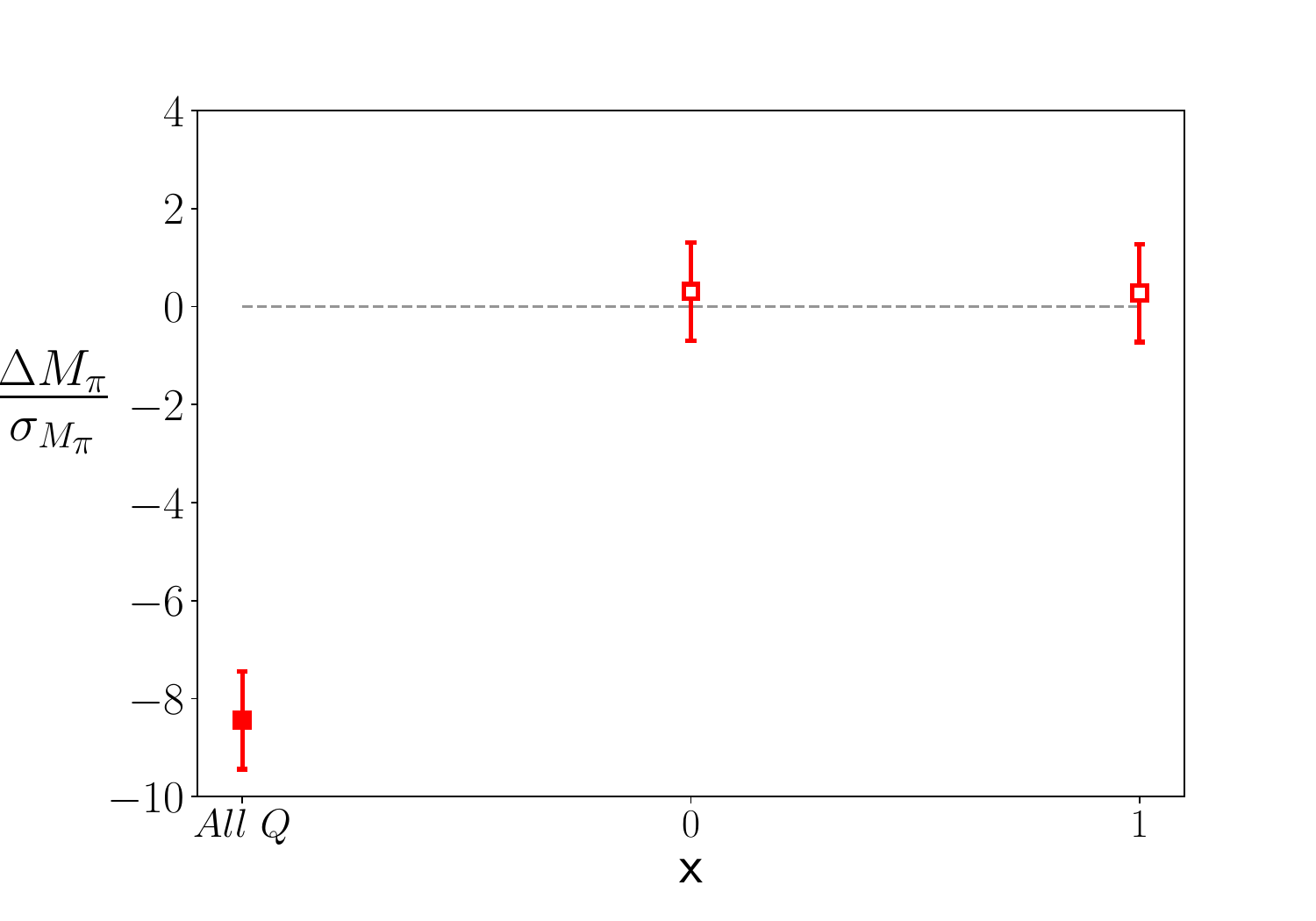}	
	\caption{ $ \Delta M_{\pi} \equiv \langle M_{\pi}\rangle_{\rm x}^{\text{HMC}} - \langle M_{\pi} \rangle
	_{\rm x}^{\text{wHMC}}$, averaged over all sectors, ${\rm x} = All\ Q$ (full symbol), or sectors of fixed ${\rm x} = |Q|$ (open symbols) as  in Eq.~(\ref{eq:defPq}),  versus $Q$ at $\beta=9.0$.}
	\label{fig:DeltaMpiQ}. 
\end{figure}

Again, Fig.~\ref{fig:wqferm} shows that HMC is not able to sample
the different topological sectors correctly at $\beta=9.0$. Focusing
on the pion mass as the observable of interest, we see
in Fig.~\ref{fig:MpiQ} that it shows a dependence on the topological
sector, explaining why HMC fails to correctly reproduce its value
in Fig.~\ref{fig:DeltaMpiQ} (label ``All $Q$'') by more than 8
standard deviations. 
Nevertheless the values of $M_\pi$ projected to the topological
sector with $|Q|=0, 1$ are correctly reproduced
(labels $|Q|=0,1$).

\section{Master-field simulations}

We now turn to the computation of physical observables by means of
simulations in large lattices, the so-called master fields~\cite{Luscher:2017cjh}. Using this
approach, observables and their errors are
computed from volume averages over a handful (even a single) of
configurations, instead of from averages over Monte 
Carlo time. Details on the determination of statistical uncertainties
using this approach are explained in appendix~\ref{app:errorMF}. 

This approach requires large volumes for reasonable error estimates
(see appendix~\ref{app:errorMF}). 
At these large values of the volume we expect the effects of the
global topology to be suppressed. 
Master-field simulations therefore bypass the effects of topology freezing 
as long as fixed topological sectors are sampled
 correctly. In section~\ref{sec:fixing-topology} we have argued that
HMC, even suffering severely from topology freezing, can
determine correctly observables on sectors of fixed
topology. Therefore we expect master-field simulations to produce
correct numbers, even if simulations are performed in a region of
parameter space where topology is frozen. 
In this section we will confirm this expectation.

We have performed simulations on lattice volumes of $8192\times 8192$
using the standard HMC algorithm at the same values of $\beta$ as in
the wHMC case (see Section~\ref{sec:puregauge}). For each case we have
generated a single configuration by a process of thermalization
(using 2000 trajectories of length 0.5), followed by an unfolding in
the two periodic directions.  
We start with a small lattice $16\times 16$. 
After unfolding 9 times, we reach our target size $8192\times 8192$. 

 \begin{figure}[h!]
	\centering
	\includegraphics[width=\linewidth,keepaspectratio]{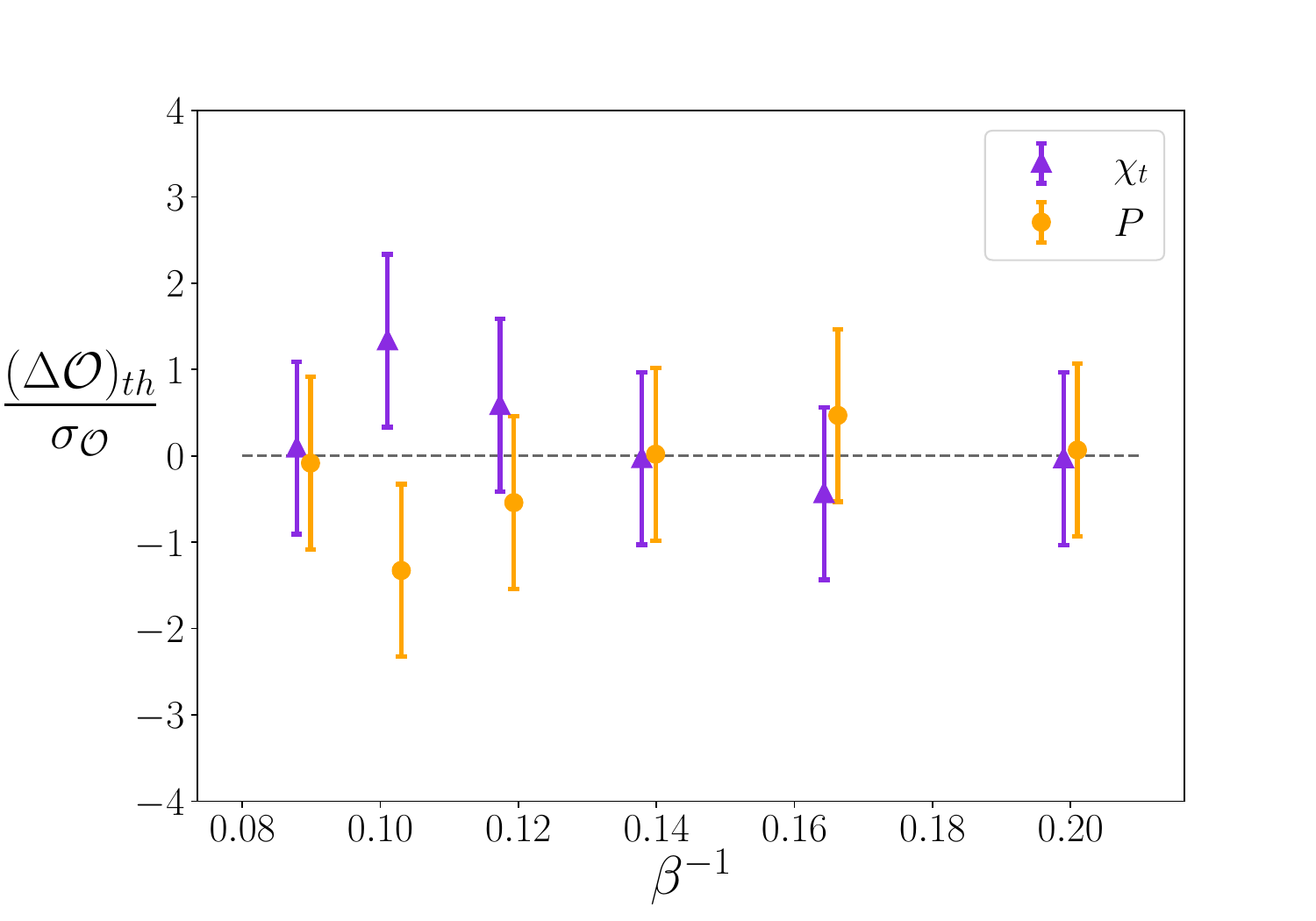}	
	\caption{ Results for the plaquette and the topological susceptibility in the master-field simulations of the pure gauge theory. We plot the difference with respect to the theoretical value $(\Delta \mathcal O)_{th}=\mathcal O - \mathcal O_{th}$, normalized by the error, as a function of $\beta^{-1}$.  }
	\label{fig:masterfield}
\end{figure}

On this single configuration, we measure the plaquette and the susceptibility. 
Given the value of the $1\times 1$ Wilson loop $U_p(x)$ at a point $x$, we use its
real part and argument to estimate the value of the plaquette and
the topological charge density respectively
\begin{equation}
  P(x) = {\rm Re}[U_p(x)]\,,\qquad q(x) = \frac{-i}{2\pi}\ln U_p(x)\,.
\end{equation}
The susceptibility can be determined from $q(x)$ using the local observable
\begin{equation}
  \label{eq:chir}
  \chi_R(x) = \sum_{x_i-y_i \leq R} q(x)q(x+y).
\end{equation}
If the value of $R$ is taken larger than the correlation
length of the system, the expectation value of $\chi_R(x)$ will
coincide with the topological susceptibility. 

In the infinite volume limit the partition function Eq.~(\ref{eq:zQ})
factorizes. 
This implies that the values of $q(x)$ are not correlated among
different $x$, and therefore $\langle \chi_t \rangle = \langle
\chi_R(x) \rangle$ for any value of $R$. 
Moreover the variables $\chi_R(x)$ are also uncorrelated.
It is easy to check that the variance of the observable
$\chi_R(x)$ increases as
\begin{equation}
  \frac{{\rm Var}[\chi_R(x)]}{{\rm Var}[\chi_0(x)]} \approx 1+2R^2\,,
\end{equation}
which implies that the best estimate of the topological
susceptibility is obtained by using $R=0$. Incidentally, this also
suggests that in theories with a non-zero correlation length $R$ has to
be taken as small as possible.

Fig.~\ref{fig:masterfield} shows that the values of the plaquette and
the susceptibility agree perfectly with the theoretical expectations. 
Further details in the evaluation of the error in master field
simulations can be found in appendix~\ref{app:errorMF}. 

Finally let us comment on the cost comparison. 
The key element for master field simulations is
the cost of thermalization. 
For our case (due to the small numerical cost of our simulations) this
thermalization has been performed by brute force. 
Whether a thermalization process performed with more care would result in a
cost comparable to the one of wHMC or HMC is beyond the scope of this work. 
Any conclusion in this regard would be anyway difficult to extrapolate
to other gauge theories in more dimensions, since this particular
model shows no spatial correlations among observables.

\section{Outlook}

We have presented a new algorithm based on Metropolis--Hastings steps
that are tailored to induce jumps in the topological charge. This
algorithm satisfies detailed balance, and ergodicity is ensured when
alternated with standard HMC steps. As we have shown, it successfully
improves the problem of topology freezing and exponentially-growing
autocorrelation times in the 2D model considered---both with and
without fermion content. The integrated autocorrelation
time of wHMC in the pure gauge case is very similar to the one
obtained in machine-learned flow-based sampling
algorithms~\cite{Kanwar:2020xzo,Albergo:2021vyo}, however without the
additional training cost.  

In spite of the shortcomings of  algorithms with topology freezing, we
have been able to confirm that averages in fixed topology sectors are
not affected, and agree in wHMC and HMC. This is seen both in the pure
gauge theory, where the analytical results are known at finite
$\beta$, as well as in the theory with fermions.  

We have finally compared the wHMC algorithm with the results by local
averages in very large lattices of size up to $L \sim 8000$. Our
results indicate that master-field simulations are satisfactory
in the controlled setup of this 2D model, since analytical results are
reproduced with very high accuracy.   

The interesting question is whether wHMC can be equally successful in
the case of other gauge theories in higher dimensions. In fact, the
winding step is trivial to extend to, for instance, a $SU(2)$ theory
in 4D. We have indeed carried out the naive implementation of wHMC in that
context, and found very poor acceptances---the ``curse'' of
dimensionality. We hope that less trivial implementations in 4D could
resolve this matter; we are currently exploring modifications of the algorithm
that incorporate the idea of normalizing flows \cite{Kanwar:2020xzo}.

\begin{acknowledgments}

We thank M. Garc\'ia P\'erez, D. Hern\'andez, C. Pena and S. Witte for
useful discussions.  We also thank D. Cascales for his contribution to
the early stages of this work. Part of this work has used a code
developed by C. Urbach~\cite{schwingerurbach}.  
 
We acknowledge support from the Generalitat Valenciana grant PROMETEO/2019/083,
the European project H2020-MSCA-ITN-2019//860881-HIDDeN, and the national
project FPA2017-85985-P. AR and FRL acknowledge financial support from
Generalitat Valenciana through the plan GenT program (CIDEGENT/2019/040). DA
acknowledges support from the Generalitat Valenciana grant ACIF/2020/011. The work
of FRL has also received funding from the EU Horizon 2020 research and
innovation program under the Marie Sk{\l}odowska-Curie grant agreement No.
713673 and La Caixa Foundation (ID 100010434).  We acknowledge the
computational resources provided by Finis Terrae II (CESGA), Lluis Vives (UV)
and Tirant III (UV).  The authors also gratefully acknowledge the computer
resources at Artemisa, funded by the European Union ERDF and Comunitat
Valenciana, as well as the technical support provided by the Instituto de Física
Corpuscular, IFIC (CSIC-UV).
 
\end{acknowledgments}

\vspace{0.5cm}

\appendix

\section{Statistical uncertainties in master-field simulations} \label{app:errorMF}

Here we present a strategy for data analysis in lattice field theory for the
case of master-field simulations, i.e., 
simulations on very large volumes, where expectation values are
determined as volume averages. 
We note that this strategy is completely analogous to the well-known
$\Gamma$-method~\cite{Madras:1988ei, Wolff:2003sm}, with invariance under
spatial translations playing a similar role as invariance under
simulation time. 

Primary observables are labeled $A_i^\alpha$, where the index $i$
labels which observable is measured on the master-field 
labeled by $\alpha$ with volume\footnote{We assume that
  master field simulations are generated with periodic boundary
  conditions on a symmetric $d$-dimensional lattice. 
  The generalization to other cases is straightforward.
} $V_\alpha = (L_\alpha)^d$.
The measurements of the observable on each point 
of the space 
\begin{equation}
  a_i^\alpha(x)\,,\qquad
  (x\in V_\alpha)\,,
\end{equation}
are used to estimate the values of the primary observables $A_i^\alpha$. 
Being precise, we use
\begin{equation}
  \bar a_i^\alpha = \frac{1}{V_\alpha} \sum_{x\in V_\alpha} a_i^\alpha(x)\,.
\end{equation}
It is also convenient to define the fluctuations over the mean,
\begin{equation}
  \delta_i^\alpha(x) = a_i^\alpha(x) - \bar a_i^\alpha\,.
\end{equation}

In general we are interested in computing the uncertainty on derived
observables. These are functions of the primary observables. 
\begin{equation}
  F\equiv f(A_i^\alpha)\,.
\end{equation}
Note that in general derived observables depend on measurements performed in
various master field simulations, possibly with different physical
parameters (lattice spacing, quark masses, volume, etc\dots).
The observable $F$ and its error are estimated by Taylor expanding around
$\bar a_i^\alpha$,
\begin{equation}
  f(a_i^\alpha(x)) = f(\bar a_i^\alpha) +
  \bar f_i^\alpha \delta_i^\alpha(x) + \dots\,,
\end{equation}
where $\bar f_i^\alpha = {\partial f}/{\partial A_i^\alpha}
\Big|_{\bar a_i^\alpha}$. 
This last equation suggests to use as estimate for the
observable 
\begin{equation}
  \bar F = f(\bar a_i^\alpha)\,,
\end{equation}
In order to compute its error, we use the autocorrelation functions
$\Gamma_F^\alpha(x)$, which can be estimated from the data using 
\begin{equation}
  \label{eq:Ferr_est}
  \Gamma_F^\alpha(x) = \frac{1}{V_\alpha}\sum_{i,j}\bar f_i^\alpha\bar f_j^\alpha
  \sum_{x'\in V_\alpha}\delta_i^\alpha(x'+x)\delta_j^\alpha(x')\,.
\end{equation}
At large distances compared with the largest correlation length in the
system $\xi_\alpha$, they decay exponentially
\begin{equation}
  \label{eq:Gdecay}
  \Gamma_F^\alpha(x) \simas{x\to\infty} e^{-|x|/\xi_\alpha}  \,.
\end{equation}
Only if $L_\alpha \gg \xi_\alpha$ it is possible to give a reasonable
estimate of the uncertainty. 
In these cases we use
\begin{equation}
  \label{eq:Ferr}
  (\delta\bar F)^2 = \sum_\alpha \frac{1}{V_\alpha}
  \sum_{x\in V_\alpha} \Gamma_F^\alpha(x)\,.
\end{equation}

In practice the summation in Eq.~(\ref{eq:Ferr}) has to be restricted to
$|x| < R$. 
As in the case of error estimation of Monte Carlo data, the optimal
value of $R$ has to be chosen as a balance between a small value, which will
underestimate the true value of the error in
Eq.~(\ref{eq:Ferr}), and a large value, which will only add
statistical noise to the error estimate. 
Similar recipes to the ones used in usual Monte Carlo simulations (see~\cite{Wolff:2003sm})
can be used to estimate appropriate values of $R$. 
Note however that in contrast with the case of Monte Carlo simulations,
the exponential asymptotic decay of the autocorrelation function in
Eq.~(\ref{eq:Gdecay}) can be estimated from the physical parameters of
the simulation. 
This opens the door to more accurate error estimates along the lines
of~\cite{Schaefer:2010hu}.  

Finally let us comment two more points. 
First, if more than one configuration is produced in a master-field
simulation, they can be used to reduce the uncertainty in the
determination of the correlation function $\Gamma_F(x)$ along the
lines of the analysis of different \emph{replica}~\cite{Wolff:2003sm}. 
Second, analyzing derived observables that depend both on master-field
simulations and Monte Carlo ensembles can be performed along the lines
suggested in~\cite{Virotta2012Critical, Ramos:2018vgu}.

\bibliography{biblio.bib}

\end{document}